\def\RELEASE{0}  %
\def\ANON{0}     %
\def\SQUEEZE{0}  %

\documentclass[sigconf,screen]{acmart}
\PassOptionsToPackage{hyphens}{url}\usepackage{hyperref}

\usepackage[noend]{algpseudocode}
\usepackage{algorithmicx}
\usepackage{booktabs}
\usepackage{caption}
\usepackage{comment}
\usepackage[inline]{enumitem}
\usepackage{graphicx}
\usepackage{listings}
\usepackage{multirow}
\usepackage{xcolor}
\usepackage{xspace}
\usepackage{libertine}
\usepackage{microtype}
\usepackage{tikz}
\usepackage{gnuplot-lua-tikz}

\usepackage{tabularx}
\usepackage{booktabs}

\usepackage{subcaption} %

\usepackage{relsize}
\usetikzlibrary{calc}

\microtypecontext{spacing=nonfrench}

\hypersetup{
  pdflang={en},
  pdfdisplaydoctitle}

\definecolor[named]{OurPurple}{cmyk}{0.55,1,0,0.15}
\definecolor[named]{OurDarkBlue}{cmyk}{1,0.58,0,0.21}
\hypersetup{
  colorlinks,
  linkcolor=OurPurple,
  citecolor=OurPurple,
  urlcolor=OurDarkBlue,
  filecolor=OurDarkBlue}

\if \SQUEEZE 1
\setlist[itemize]{
  leftmargin=*,
  itemsep=2pt,
  topsep=2pt}

\fi

\usepackage{color}
\definecolor{deepblue}{rgb}{0,0,0.5}
\definecolor{deepred}{rgb}{0.6,0,0}
\definecolor{deepgreen}{rgb}{0,0.5,0}

\def\Snospace~{\S{}}

\newcommand{\circled}[1]{{\small\protect\raisebox{0.5pt}{\textcircled{\raisebox{-.2pt}{\textls[-50]{\relsize{-1.5}\phantom{0}\makebox[0pt][c]{#1}\phantom{0}}}}}}\xspace}

\if \RELEASE 1
  \def\NOTES{0}
\else
  \def\NOTES{1}
\fi
\if \NOTES 1
  \newcommand{\XXX}[1]{{\color{red}{XXX {#1}}}}
  \newcommand{\antoine}[1]{{\color{teal}{[\textbf{AK:} {#1}]}}}
  \newcommand{\todo}[1]{{\color{blue}{TODO: {#1}}}}
\else
  \newcommand{\XXX}[1]{}
  \newcommand{\antoine}[1]{}
  \newcommand{\todo}[1]{}
\fi

\newcommand{\eg}{e.g.\xspace}

\if \ANON 1
  \newcommand{\sys}{Contour\xspace}
  \newcommand{\spectroviz}{SpectroViz\xspace}
  \newcommand{\parallax}{Trace\emph{ll}ens\xspace}
  \newcommand{\tracecomp}{TraceComp\xspace}
  \newcommand{\tracecompx}{tracecomp}
\else
  \newcommand{\sys}{Contrast\xspace}
  \newcommand{\spectroviz}{SpectroViz\xspace}
  \newcommand{\parallax}{Para\emph{ll}ax\xspace}
  \newcommand{\tracecomp}{TraceComp\xspace}
  \newcommand{\tracecompx}{tracecomp}
\fi

\newcommand{\codefrag}[1]{{\ttfamily\small #1}}

\lstdefinelanguage{Go}{
  keywords={break, case, chan, const, continue, default, defer, else, fallthrough, 
            for, func, go, goto, if, import, interface, map, package, range, return, 
            select, struct, switch, type, var},
  keywordstyle=\color{blue}\bfseries,
  ndkeywords={bool, byte, complex64, complex128, error, float32, float64, 
              int, int8, int16, int32, int64, rune, string, uint, uint8, 
              uint16, uint32, uint64, uintptr},
  ndkeywordstyle=\color{blue}\bfseries,
  identifierstyle=\color{black},
  comment=[l]{//},
  morecomment=[s]{/*}{*/},
  commentstyle=\color{deepgreen}\ttfamily,
  stringstyle=\color{red}\ttfamily,
  sensitive=true,
  morestring=[b]",
  morestring=[b]`,
}

\lstdefinestyle{golang}{
    basicstyle=\scriptsize\ttfamily,
    breakatwhitespace=false,      
    breaklines=true,                 
    captionpos=b,                    
    keepspaces=true,                 
    numbers=left,                    
    numbersep=5pt,                  
    showspaces=false,                
    showstringspaces=false,
    showtabs=false,                  
    tabsize=1,
    xleftmargin=1.5em,
    frame=single,
    framexleftmargin=1.5em,
}

\usepackage[most]{tcolorbox}
\tcbuselibrary{xparse,listings}
\NewTCBListing{framedlisting}{O{}}
 {%
  boxrule=0.4pt,
  colback=white,
  colframe=black,
  sharp corners,
  listing only,
  left=0mm, right=0mm, top=0mm, bottom=0mm,
  listing options={#1,frame=none, xleftmargin=0em, framexleftmargin=0em, framextopmargin=0em, belowskip=0em, aboveskip=0em},
 }

\newcommand{\fakepara}[1]{\noindent\textbf{#1}\xspace}

\newcommand{\evalgraphscale}{0.25}

\newcounter{mysubsub}[subsection]
\newcommand{\mysubsubsection}[1]{%
  \refstepcounter{mysubsub}%
  \vspace{0.25\baselineskip}
  \noindent\textbf{\thesubsection.\themysubsub.\ #1}\par
  \vspace{0.25\baselineskip}
}

\setcopyright{none}
\renewcommand\footnotetextcopyrightpermission[1]{} %

\if \ANON 1
\acmConference[SoCC'26]{SoCC'26}{Nov 2026}{Singapore}
\fi

\begin{document}
\date{}
\title{Enabling Multi-Dimensional Distributed Trace Comparison with \sys}
\if \ANON 1
\subtitle{Full Research Paper \#231 (\pageref{page:last} pages)}
\fi

\if \ANON 1
  \author{Full Research Paper \#231 (\pageref{page:last pages})}
\else
  \author{Vaastav Anand}
  \affiliation{
    \institution{Max Planck Institute for Software Systems}
    \country{}
  }
  \author{Rodrigo Fonseca}
  \affiliation{
    \institution{Microsoft Azure Research}
    \country{}
  }
  \author{Jonathan Mace}
  \affiliation{
    \institution{Microsoft Research}
    \country{}
  }
  \author{Antoine Kaufmann}
  \affiliation{
    \institution{Max Planck Institute for Software Systems}
    \country{}
  }
\fi

\begin{abstract}
Distributed traces are widely used for diagnosing performance anomalies in modern cloud applications. 
However, diagnosis is fundamentally a comparative task: operators seek to understand how an anomalous execution differs from expected behavior, 
how a deployment changes system execution, or how two individual executions differ. 
Trace comparison is challenging because useful differences between executions can manifest across multiple dimensions, 
and no single diagnostic interface is effective at capturing all of them. 
Moreover, the relevant dimensions and comparison populations are often not known a priori; 
operators construct and refine comparison sets dynamically as they develop hypotheses about system behavior. 
Consequently, effective trace comparison requires flexible representations 
that preserve diverse execution characteristics while supporting multiple diagnostic perspectives.

This paper presents \sys, a system for multi-dimensional comparative trace analysis. 
\sys introduces the Trace Projection Object (TPO), a mergeable representation that captures structural, temporal, critical-path, and semantic 
properties of trace populations while enabling efficient construction of arbitrary comparison sets at query time. 
Unlike approaches that define a fixed notion of trace difference, \sys separates trace representation from comparison semantics, 
allowing diverse interfaces to selectively reason about specific dimensions.
This separation enables the composition of complementary interfaces, 
allowing operators to combine insights from multiple dimensions for more effective diagnosis. 
We demonstrate this capability through two complementary interfaces: 
(i) \spectroviz, a critical-path-based visual interface for localizing execution differences; and 
(ii) \parallax, a natural language interface for generating explanations of trace differences using LLMs.
  
Through \sys, we establish trace comparison as a first-class abstraction for distributed system diagnosis 
and provide a foundation for building diverse comparison interfaces over a shared representation. 
We demonstrate the effectiveness and efficiency of \sys through controlled experiments on traces
from DeathStarBench and evaluation on production traces from Uber.
\end{abstract}
 
\maketitle

\section{Introduction}

Distributed tracing~\cite{sigelman2010dapper,barham2004using} is a major pillar of modern cloud observability~\cite{sridharan2018distributed,opentelemetry}.
A distributed trace captures the execution structure, timing, and semantics of an individual request as it traverses
through the various components of a system. 
These traces provide an execution-centric view of system behavior that is indispensable for performance diagnosis, debugging, and root-cause analysis. 

Despite their widespread adoption~\cite{mace2017end}, traces remain difficult to interpret in practice~\cite{davidson2023qualitative}. 
A single request may span hundreds of operations across numerous services, while production systems generate millions of traces daily~\cite{huye2023lifting,kaldor2017canopy}. 
Moreover, traces contain dense information about the request execution across multiple dimensions: temporal execution behavior,
causal structural behavior, error signals and stack traces, unstructured logging statements, and developer-annotated tags
and attributes.
Consequently, developers need to analyse and explore the various dimensions of large traces for
carrying out diagnostics tasks.

For consumers of tracing, the most important task is trace comparison
as operators rarely reason about traces in isolation. 
Instead, diagnosis is inherently a comparative process: operators seek to understand why one execution differs from another, why a particular request 
deviates from expected behavior, or how system behavior changes across workloads, deployments, and time.
Across a wide range of diagnostic workflows, we observe three recurring comparison tasks: 
(i) comparing two individual traces, 
(ii) comparing an anomalous trace against a population of normal traces, and 
(iii) comparing two populations of traces. 

Executing trace comparison tasks is difficult today for a multitude of reasons.
First, existing tracing tools provide only limited support for comparative analysis.
Most systems present individual traces through timeline or graph visualizations, leaving users to manually identify differences across executions.
Second, existing systems with comparison interfaces typically co-design the processing of selected dimensions with the interface itself. 
Consequently, they are optimized for highlighting differences along a limited set of dimensions and may discard potentially useful information, 
as no single interface is equally effective for capturing and presenting all dimensions of trace behavior.
Finally, the exact useful populations involved for trace comparison tasks are rarely known a priori. 
Operators dynamically construct comparison populations based on evolving hypotheses, 
filtering traces by service, deployment version, latency range, error type, or workload. 

As a result, operators must manually combine multiple tools and representations to answer fundamental diagnostic questions.
To aid developers in answering bespoke trace comparison queries, a trace comparison system must fulfill the following requirements.
First, supporting exploratory workflows requires a representation that enables efficient construction and comparison of arbitrary trace populations 
without repeatedly operating on raw traces. 
Second, the trace comparison system must support the development of complementary comparison interfaces that can operate on a common substrate
of raw trace data as different diagnostic interfaces emphasize different aspects of execution behavior.

This paper introduces \sys, a framework that natively supports design and implementation of comparison interfaces
for the different comparison tasks on dynamic trace populations. To achieve this, \sys introduces the Trace Projection Object (TPO), a canonical and mergeable representation for comparative trace analysis. A TPO captures 
structural, temporal, semantic, and critical-path characteristics of executions while supporting efficient aggregation over dynamically selected trace populations. 
Rather than storing pre-defined populations or fixed baselines, the system stores trace-level projections and constructs population-level TPOs on demand through aggregation and merge operations. 
This design enables arbitrary population construction while preserving the information necessary for comparative reasoning.
TPO provides a common substrate for trace populations, which can then further be used by different diagnostic comparison interfaces.

\sys provides comparison interfaces that utilize TPOs to support trace diagnosis. 
We instantiate this capability through two complementary interfaces. 
The first, \spectroviz, is a visual comparison interface that localizes structural and temporal differences across the critical paths of individual traces and trace populations. 
The second, \parallax, generates natural-language explanations of observed differences using an LLM-based multi-agent system.

Together, these contributions establish a new foundation for comparative observability. By elevating trace comparison to a first-class abstraction, we enable dynamic population analysis, multi-dimensional diagnostic interfaces, and principled evaluation of trace-comparison systems.

We evaluate the effectiveness of TPOs and our comparison interfaces in both synthetic and production settings. 
To cover a broad range of trace comparison scenarios, we develop a trace comparison benchmark that enables users to introduce controlled perturbations 
into systems and generate anomalous traces. 
We apply this benchmark to trace datasets containing anomalous traces from a popular open-source microservice benchmark, 
DeathStarBench~\cite{gan2019open}, as well as a production trace dataset released by Uber~\cite{zhang2022crisp_artifact}. 
Powered by TPOs and the comparison interfaces, \sys efficiently compares anomalous traces against normal traces, 
enabling users to quickly identify critical deviations.

In summary, this paper makes the following contributions:

\begin{itemize}
  \item We characterize three common comparison tasks spanning traces and trace populations (\autoref{sec:tasks}).
  \item We introduce the Trace Projection Object (TPO), a mergeable representation that supports dynamic population construction and comparison across structural, temporal, semantic, and critical-path dimensions (\autoref{sec:design}).
  \item We present two complementary comparison interfaces that operate on TPOs: \spectroviz (\autoref{sec:spectroviz}) and \parallax (\autoref{sec:parallax}).
\end{itemize}
\section{Background and Motivation}%
\label{sec:bg}

\subsection{Primer on Distributed Tracing}

Developers and operators use distributed tracing as a critical monitoring component in modern cloud systems.
Distributed tracing provides support to troubleshoot incidents, root-cause issues, and post-mortem analysis.
To do so, distributed tracing frameworks generate an execution trace of each request.

\fakepara{Trace Structure.} A distributed trace represents the execution of one request across the various components of a system.
Each trace is rich in structural, temporal, and semantic data.
Each trace is a Directed Acyclic Graph (DAG) of spans. A span typically represents
a unit of work done by a component in the system. The granularity of what constitutes as a unit of work
is developer-defined and often varies from not only one system to another but also from one
service to another in a given system. Developers can optionally add events, typically unstructured log messages,
to the span as well as attach informative tags to the span. 
Each span records its timing and duration, as well as arbitrary key-value annotations provided by a developer.
Edges between spans represent happened-before relationships~\cite{lamport2019time} and represent communication between two components
through local function call, remote procedure calls, or other forms of communication.
Each individual trace can be very large, comprising thousands of spans and events~\cite{huye2023lifting,kaldor2017canopy,zhang2022crisp,lee2024tale}.
Typically, large-scale production systems capture traces for millions of requests per day~\cite{kaldor2017canopy}.

\fakepara{Trace Analytics Dashboards.} 
Modern tracing systems, such as Jaeger~\cite{jaeger}, provide dashboards for querying and filtering traces, service dependency graphs, and aggregate latency statistics. 
The most popular visualization used today is the Gantt chart visualization~\cite{sridharantraceviewwrong,davidson2023qualitative}. 
The Gantt chart visualization provides a timeline view of a single request execution,
enabling users to inspect the temporal relationships between spans within a request. 
Consequently, these dashboards are effective for inspecting individual executions, 
however, the main drawback of these dashboards
is their focus on individual requests. 
The prevalent Gantt Chart visualization provides no
context for whether the trace represents normal or outlier
behavior --- a feature dependent on other traces.
This lack of context makes Gantt charts inefficient for
aggregate analyses.

\subsection{Canonical Trace Comparison Tasks}
\label{sec:tasks}

Diagnosis through distributed tracing is fundamentally a comparative activity. 
Operators rarely seek to understand a trace in isolation; instead, they seek to understand how an execution differs from some expected or alternative behavior. 
Across debugging, performance analysis, deployment validation, and incident response workflows, 
we observe three recurring trace comparison tasks that form the foundation of diagnostic reasoning.
These tasks differ in the granularity of the objects being compared but share a common objective: identifying and explaining behavioral differences across executions. We discuss these tasks below in detail.

\mysubsubsection{Pairwise Trace Comparison}

The simplest task compares two individual traces. 
Such comparisons arise when operators compare a failing request against a successful request, investigate regressions between two executions, or analyze the effects of different inputs on system behavior.
For example, an operator may compare a slow request against a fast request to determine why one execution experienced elevated latency. 
The comparison may reveal that a database operation became part of the critical path, that an additional service invocation occurred, or that latency increased within a particular span.

Pairwise trace comparison is often most effective for understanding specific execution instances
as individual traces provide limited statistical context. 
A difference observed between two traces may reflect a genuine anomaly or simply natural execution variability. 
Consequently, pairwise trace comparison is incapable of establishing insights about broader system behavior.

\fakepara{Existing Techniques.} Traditional pairwise trace comparison techniques often transform the rich structural and temporal information 
contained in distributed traces into representations that are more amenable to automated comparison. 
Examples include linearizing execution paths and comparing them using string-edit distance~\cite{barham2004using,rabo2020distributed} 
or modeling execution paths using probabilistic context-free grammars~\cite{d2024grammar,chen2004path}.
However, these techniques often lose parallelism and concurrency
information, which may be detrimental for certain analysis.
To better preserve execution structure, other techniques compare traces directly as graphs.
For example, Jaeger provides a comparative view~\cite{farro2018jaegertracecompare} that 
visualizes the difference of the raw execution graphs of the two traces,
while Mace and Fonseca~\cite{mace2013revisiting} compare execution graphs using Weisfeiler-Lehman graph kernels to quantify structural similarity. However, these approaches usually do not scale well
with large graphs as they overload the visual channels of an operator~\cite{anand2020aggregate}.

\mysubsubsection{Trace-to-Population Comparison}

Many diagnostic tasks require understanding how a particular execution differs from a larger set of executions. 
In these scenarios, operators compare an individual trace against a population representing expected behavior. 
The population may consist of traces matching a particular workload, deployment version, service endpoint, or time period.
For example, an operator investigating a slow request may wish to compare the request against all successful requests to the same endpoint. 
The resulting comparison can identify which portions of the execution deviate from typical behavior.
Trace-to-population comparison combines the contextual richness of individual executions with the statistical grounding of aggregate behavior. 

\fakepara{Existing Techniques.} To support reasoning this comparison task, 
recent techniques incorporate aggregate information directly into trace visualizations~\cite{silva2021mu,traini2024vamp,anand2020aggregate,samanta2024visualizing,sambasivan2013visualizing}.
Canopy~\cite{kaldor2017canopy} can generate simple graphs of
derived metrics from traces, while Dynatrace's Service Flow feature~\cite{dynatraceserviceflow} presents aggregate request workflows and associated characteristics to facilitate high-level behavioral analysis.
Uber's trace graph diff visualization~\cite{shkurographdiffviz} compares incoming traces against a known-good population to identify anomalous executions and visualizes structural differences as a directed graph. 
TraVista~\cite{anand2020aggregate} augments the single trace Gantt-chart visualization with
histograms showing the latency distributions of different spans 
in the trace and additionally highlights infrequent RPC calls between services. 
While these systems demonstrate the value of combining execution-level and aggregate information, 
they assume a fixed aggregate computed over a predetermined trace set. 
Consequently, operators cannot dynamically construct comparison populations as diagnostic hypotheses evolve. 
Indeed, the maintainers of Jaeger identified support for dynamically selected comparison populations as a prerequisite for integrating TraVista's 
visualizations into Jaeger's trace viewer~\cite{jaegertravistaissue}. 

\mysubsubsection{Population-to-Population Comparison}

The most general comparison task involves comparing two populations of traces. 
Such comparisons arise when operators seek to understand behavioral changes across deployments, workloads, configurations, geographic regions, or time periods.
Population-to-population comparisons enable the discovery of systemic behavioral shifts that may not be visible from individual traces. 
For instance, a deployment may introduce a new service dependency, alter critical-path composition, increase latency variance within a subset of services, or change the frequency of specific execution paths. 
These effects may only become apparent when aggregated across many executions.
Unlike trace-to-trace comparisons, population-to-population comparisons emphasize distributional changes rather than individual execution differences. 
The goal is not to identify what happened in a particular request, but rather to characterize how overall system behavior has evolved.

\fakepara{Existing Techniques.}
Early approaches, such as Spectroscope~\cite{sambasivan2007categorizing,sambasivan2011diagnosing} compared latency distributions across trace populations but required the compared traces to be structurally isomorphic, 
an assumption that often does not hold in modern distributed systems where traces may exhibit diverse execution paths~\cite{las2019sifter}. 
Pintrace~\cite{pintrace} enables users to select and compare metric distributions across two trace populations, 
DTraComp~\cite{ekhlasi2026dtracomp} further explores interactive population comparison through color-coded differential frame graphs. 
While these systems demonstrate the utility of population-level comparison, they typically couple the comparison logic with a specific visualization 
or analysis technique, limiting their extensibility to other dimensions that may carry relevant signals. 

\subsection{Challenges}

Building a general-purpose trace comparison system presents three key challenges.

\fakepara{Challenge 1: Dynamic Comparison Populations.}
Unlike traditional observability queries, trace comparison rarely operates over fixed datasets. 
Operators iteratively construct comparison populations as they refine diagnostic hypotheses, filtering traces by workload, deployment version, service, latency range, error type, or structural characteristics~\cite{davidson2023qualitative}. 
Consequently, the populations involved in a comparison are often unknown until query time.
A straightforward approach would repeatedly scan and analyze raw traces whenever a new population is selected. 
However, such an approach becomes prohibitively expensive as trace volumes grow and comparison populations evolve interactively. 
The challenge is to support efficient construction and comparison of arbitrary trace populations while preserving the structural, temporal, and semantic information necessary for diagnosis.

\fakepara{Challenge 2: No Single Interface Captures All Diagnostic Dimensions.}
Distributed system behavior manifests across multiple dimensions, including execution structure, temporal behavior, critical paths, and semantic changes. 
However, these dimensions are best understood through different diagnostic interfaces. 
For example, a visualization may be effective for localizing where behavior changed within an execution, 
while a language-based explanation may be better suited for summarizing high-level causal differences between trace populations. 
Similarly, a topology-oriented interface may reveal structural divergence, whereas a performance-oriented interface may highlight latency shifts and critical-path changes.

Consequently, a single comparison interface cannot effectively capture and communicate all relevant aspects of trace differences. 
Interfaces must make trade-offs in the dimensions they emphasize, potentially obscuring other useful information. 
Moreover, the dimensions most relevant for diagnosis are often not known a priori and depend on the operator's investigative goal.

Therefore, an effective trace comparison system requires a representation that preserves multiple dimensions of execution behavior independently of any specific interface. 
Such a representation enables different diagnostic modalities to selectively project and compare the dimensions most relevant to their intended use case.

\fakepara{Challenge 3: Bridging Population-Level Patterns and Individual Executions.}
Diagnosis requires reasoning at multiple levels of aggregation. 
Aggregate comparisons reveal population-level behavioral shifts, but operators ultimately investigate concrete executions. 
For example, identifying that database operations contribute disproportionately to a latency regression is useful, 
but operators must still locate the corresponding spans within specific traces to understand why the regression occurred.
The challenge is to enable seamless navigation between population-level comparisons and the raw traces that give rise to them.

These challenges motivate a representation that supports dynamic population construction, enables multiple comparison semantics, and remains grounded in concrete executions. %
\section{\sys Design}%
\label{sec:design}

\if \ANON 1
\begin{figure*}%
\centering%
\includegraphics[width=\linewidth]{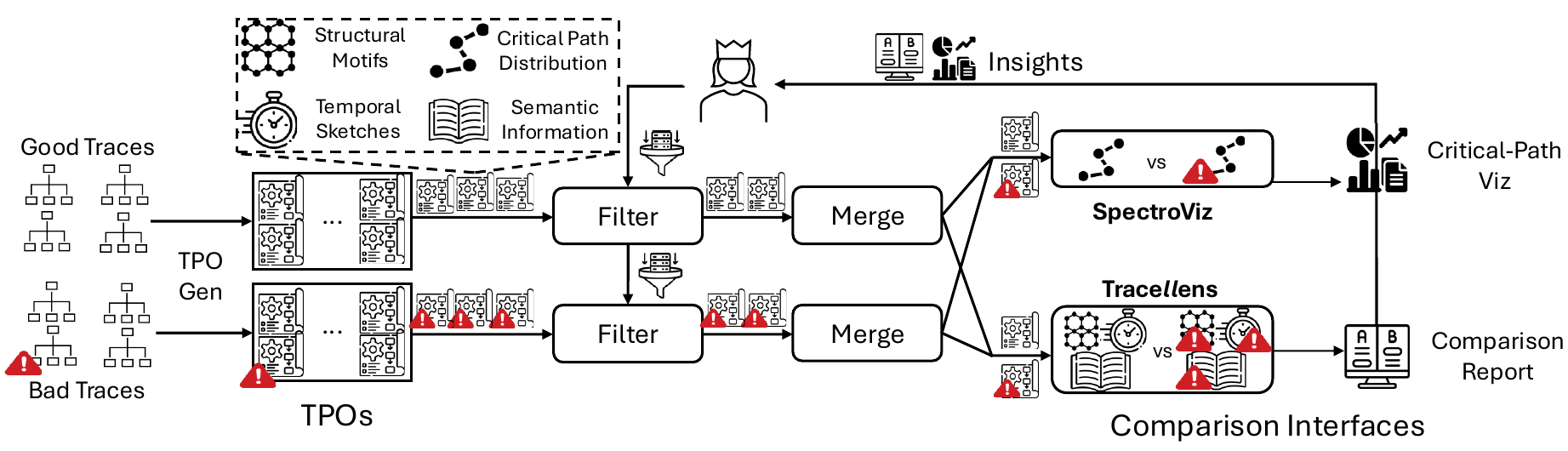}%
\caption{\sys Workflow}%
\label{fig:contrast_workflow}
\end{figure*}
\else
\begin{figure*}%
\centering%
\includegraphics[width=\linewidth]{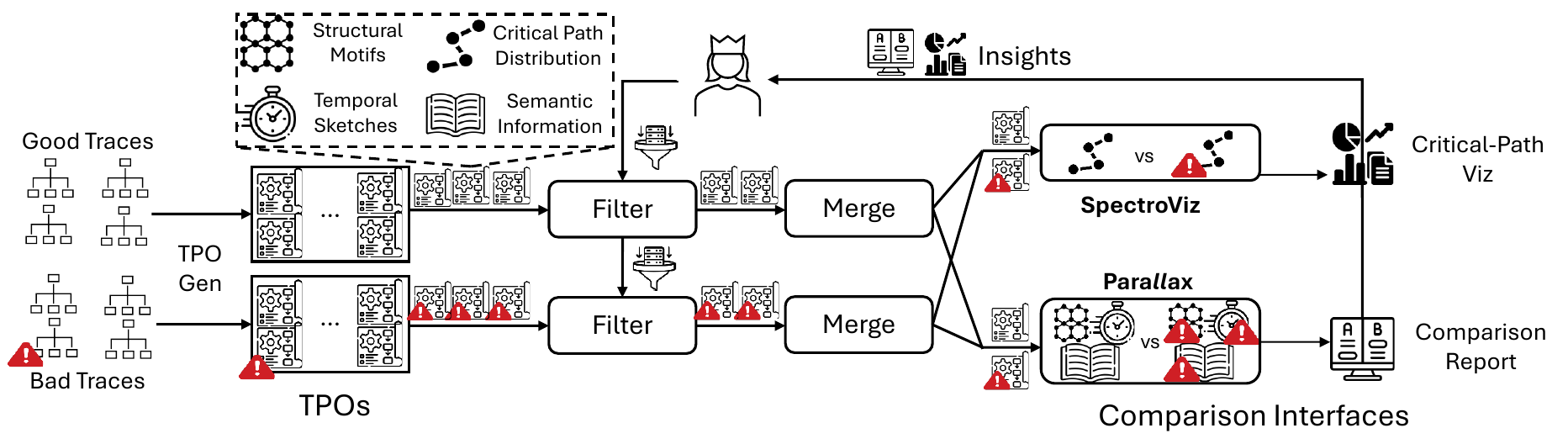}%
\caption{\sys Workflow}%
\label{fig:contrast_workflow}
\end{figure*}
\fi

To address these challenges, we introduce \sys,
a system for enabling iterative trace comparison tasks.
The underlying contribution of \sys is the Trace Projection Object (TPO), 
a canonical representation that captures the comparison-relevant characteristics of traces while supporting efficient aggregation 
and dynamic population construction.

\autoref{fig:contrast_workflow} illustrates the workflow enabled by \sys. 
Given two input trace populations,a population containing predominantly healthy traces and another containing anomalous or faulty traces,
\sys first processes each trace to extract high-level attributes, such as start time, end-to-end latency, error tags, and operations, which enable flexible trace filtering. 
It then generates an individual TPO for each trace. 
Users can optionally apply filters over these attributes to refine the trace populations, 
after which \sys aggregates the remaining trace-level TPOs into a single population-level TPO for each population. 
\sys then provides the resulting good and anomalous population TPOs to its comparison interfaces. 
Each interface leverages different dimensions of the TPO representation to compare the populations and generate human-consumable insights. 
Based on these insights, users can iteratively refine their filters and repeat the comparison process to progressively narrow down the root causes of anomalous behavior.

\subsection{Trace Projection Object}

The Trace Projection Object (TPO) is the underlying data model that captures the information required for comparative diagnosis tasks. 
We design the TPO around three key requirements. 
First, it supports the three canonical comparison tasks: trace-to-trace, trace-to-population, and population-to-population comparison. 
Second, it enables dynamic population construction by supporting efficient aggregation and composition of trace representations without requiring repeated access to raw traces. 
Third, it remains independent of any specific comparison modality. Visual analytics, language-based explanations, and future diagnostic interfaces can 
operate over the same underlying representation while defining their own mechanisms for extracting and presenting comparative insights.
The TPO extracts four classes of information that capture the comparison-relevant behavior of the execution.

\fakepara{Structural Information.} The structural information parses the execution graph of the trace
to extract structural motifs from the trace. The TPO creates a mapping of
the unique motifs that appeared in the trace along with the frequency
with which they appeared.
Extracting the motifs once at TPO generation time avoids re-processing of raw traces
to extract different structural information.
Currently, \sys captures unique parent-child relationships from the trace as part of the TPO.
However, this can be further extended to capture more prominent workflow motifs~\cite{abdi2025workflow}.

\fakepara{Temporal Information.} The temporal component captures the execution timing of the trace.
For each unique operation in a trace, TPO creates a latency distribution sketch using the DDSketch algorithm~\cite{masson2019ddsketch}.
This allows TPOs to effectively capture latency distributions for an operation that appears multiple times in a trace.
While using a sketch may seem counterintuitive for every unique operation in a single trace,
it allows TPOs to be easily mergeable as merging two sketches is a much cheaper operation than
merging raw arrays of latency distributions.

\fakepara{Critical-Path Information.} The critical-path component captures the execution segments responsible for end-to-end latency~\cite{lightstepcritical}. 
\sys identifies these segments using Jaeger's modified implementation of the CRISP Critical Path algorithm~\cite{zhang2022crisp}. 
Rather than representing the critical path as a sequence of entire spans, \sys decomposes it into finer-grained span sub-segments. 
For example, if a database query overlaps with multiple downstream RPCs, only the portion of the query that is not hidden by concurrent execution lies on the critical path. 
By recording these sub-segments, \sys more precisely localizes the execution intervals that contribute to overall request latency. 
To store this, TPO maintains a one-dimensional occupancy array in the TPO for each unique operation,
where each array index corresponds to a discrete time interval, 
and the value at that index indicates whether the operation was executing on the critical path during the corresponding interval.

\fakepara{Semantic Information.} The semantic component captures high-level execution properties that are not directly reflected in a trace's structural or temporal characteristics, yet provide important context for diagnosis. 
Unlike execution topology or latency measurements, semantic information describes \emph{what} the application was doing and \emph{why} it behaved in a particular way. 
As part of the TPO, \sys records the set of unique log statements associated with each operation, along with their occurrence frequencies. 
It also maintains frequency counters for observed request attributes, including errors, error types, tags, and other key-value attributes.

\subsection{TPOs for Trace Populations}

\fakepara{TPO Merge Operator.} Individual trace projections serve as the building blocks for population-level representations. 
Given a trace population, $P(q)={t_1,t_2,\dots,t_n}$ defined by a user query $q$, \sys constructs a population TPO by repeatedly applying a merge operator to the TPOs of individual traces. 
$
\text{Merge}
(
\text{TPO}(P_1),
\text{TPO}(P_2)
)
\rightarrow
\text{TPO}(P_1 \cup P_2)
$
More generally, the same operator can merge any two TPOs, where an individual trace is treated as a population of size one. 
Thus, constructing a TPO for an entire trace population corresponds to repeatedly merging the TPOs of individual traces.

The Merge operator independently combines each dimension of the representation. 
For structural information, it adds the frequency counters associated with each unique structural relationship, 
producing a combined view of execution patterns across both populations. 
For temporal information, it merges the DDSketch~\cite{masson2019ddsketch} quantile sketches associated with each unique operation, preserving the aggregated latency distributions. 
For critical-path information, it performs an element-wise addition of the occupancy arrays for each operation, 
accumulating the number of traces executing that operation on the critical path during each time interval. 
Finally, for semantic information, it unions the sets of unique log statements associated with each operation while adding their occurrence counts, 
and similarly adds the frequency counters maintained for request attributes, errors, tags, and other key-value attributes.

\fakepara{Representative Trace Context.}
While population-level projections enable scalable comparison, operators ultimately reason about individual executions. 
Consequently, \sys does not treat TPOs as replacements for raw traces.
Instead, each TPO maintains references to representative traces drawn from the underlying population. 
Representative traces may include average-case executions, highly divergent traces, critical-path exemplars, or user-selected traces.
These traces provide execution-level context that can be used by comparison interfaces to ground aggregate observations in concrete executions.
\section{Comparison Interfaces for TPOs}%
\label{sec:impl}

TPO does not define a single notion of difference
as different diagnostic interfaces require different comparison semantics. 
For example, a visual renderer may emphasize temporal localization of execution changes, 
while a language-based renderer may focus on causal explanations and critical-path shifts.
Consequently, the role of the TPO is not to compute differences directly, 
but rather to provide a common substrate from which multiple comparison modalities can derive their own task-specific notions of divergence.
This separation between representation and comparison enables \sys to support diverse diagnostic interfaces 
with a unified underlying data model.

\fakepara{Complementary Comparison Interfaces for TPOs.} No single comparison interface is 
equally effective at exposing differences across all dimensions of a trace. 
Each interface naturally emphasizes a particular subset of the information captured by the TPO.
Rather than seeking a universal comparison interface, \sys embraces complementary diagnostic modalities that operate over the same underlying TPO. 
By presenting the same trace representation through multiple comparison interfaces, 
operators can leverage the strengths of each modality while compensating for their individual limitations, 
resulting in a more complete understanding of behavioral differences.

To demonstrate the generality of TPOs, we design two such comparison interfaces.
First, \spectroviz is a visualization comparative interface that uses the critical path dimension of TPOs to support pairwise trace and trace-to-population comparisons.
Second, \parallax is an LLM-based multi-agent system that primarily uses the semantic, temporal, and structural 
dimensions to generate a natural language explanation for TPO comparisons.
Together, these two comparative interfaces illustrate how fundamentally different diagnostic modalities 
can be built on top of the same underlying trace representation.

\subsection{\spectroviz}
\label{sec:spectroviz}

\begin{figure}[t]%
\centering%
\begin{tikzpicture}
\node[anchor=south west, inner sep=0] (img) at (0,0)
  {\includegraphics[width=\linewidth]{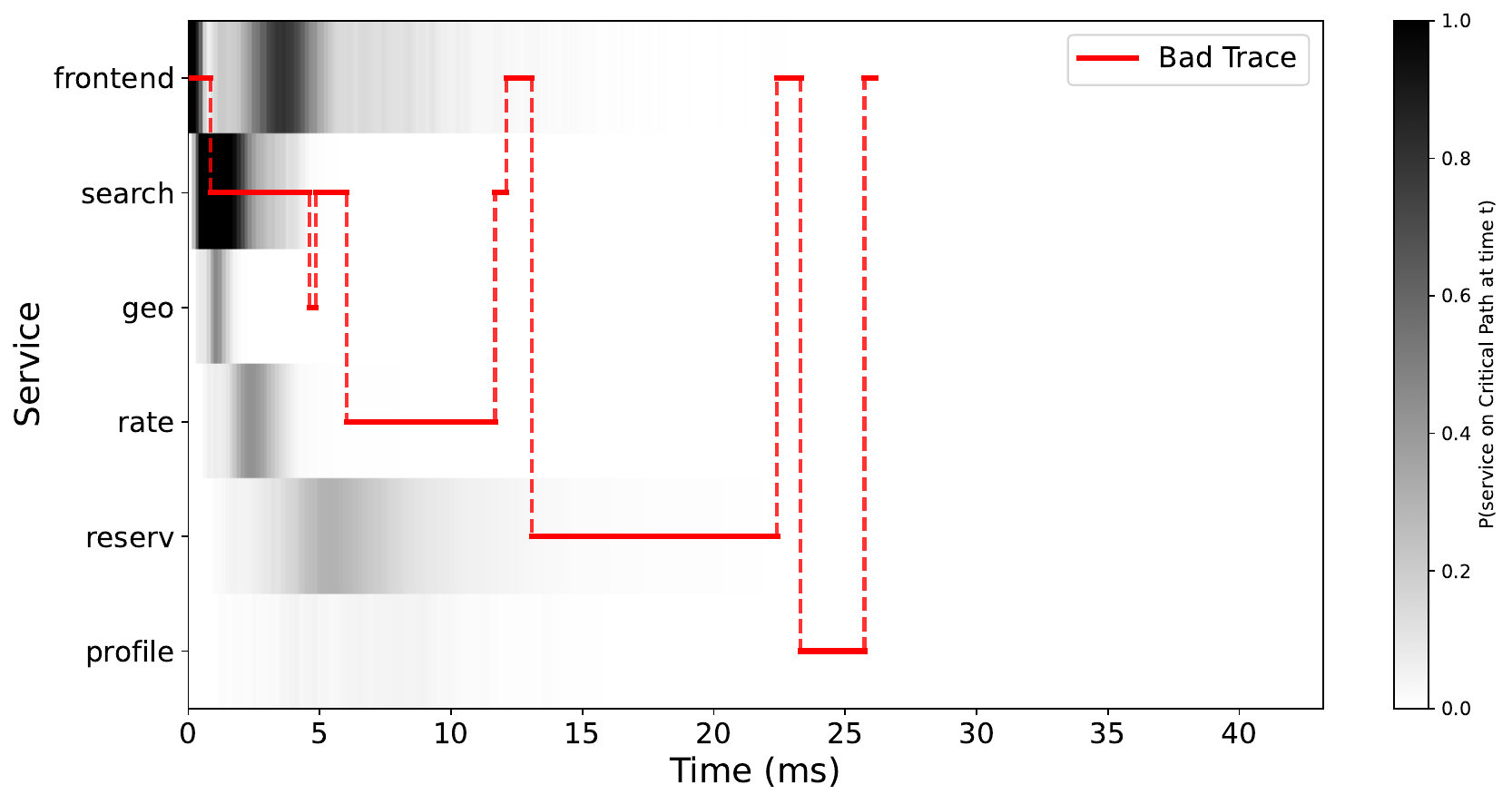}};

    \begin{scope}[x={(img.south east)}, y={(img.north west)}]

        \node[draw,circle,fill=white,thick,minimum size=1mm,font=\small]
            at (0.40,0.05) {1};

        \node[draw,circle,fill=white,thick,minimum size=1mm,font=\small]
            at (0.02,0.35) {2};

        \node[draw,circle,fill=white,thick,minimum size=1mm,font=\small]
            at (0.91,0.20) {3};

        \node[draw,circle,fill=white,thick,minimum size=1mm,font=\small]
            at (0.25,0.90) {4};

        \node[draw,circle,fill=white,thick,minimum size=1mm,font=\small]
            at (0.25,0.19) {5};

        \node[draw,circle,fill=white,thick,minimum size=1mm,font=\small]
            at (0.43,0.40) {6};
        
        \node[draw,circle,fill=white,thick,minimum size=1mm,font=\small]
            at (0.62,0.67) {7};

        \node[draw,circle,fill=white,thick,minimum size=1mm,font=\small]
            at (0.27,0.75) {8};

    \end{scope}
\end{tikzpicture}
\caption{\spectroviz annotated example for a temporal perturbation trace from the benchmark compared to a population of traces with no perturbations}%
\label{fig:spectroviz_annotated}%
\end{figure}

We present a visualization interface for TPOs called \spectroviz for pairwise trace and trace-to-population comparison tasks.
To assist in these comparison tasks, \spectroviz uses the critical path dimension of the TPO of an anomalous
trace and compares it with the critical path dimension of the TPO of a set of baseline traces.
\spectroviz generates a single visualization where it first plots the occupancy probability distribution of the critical path at each API (or service)
from the baseline population of traces and then overlays the critical path of the anomalous trace on top of the occupancy distribution.
With this visualization, operators can easily spot the difference in the critical path behavior of the anomalous
trace from its expected behavior.

We explain the visualization specifics and underlying theory using the annotated example of a trace-to-population comparison shown in \autoref{fig:spectroviz_annotated}.
In the figure, \spectroviz generates a visualization that compares a bad trace and
a good population of traces for the Search API exposed by the Frontend Service of Blueprint's implementation of Hotel Reservation System from the DeathStarBench microservice benchmark.
We will refer to the circled numbers in the figure (\eg \circled{1}) throughout this section.

\fakepara{Interface Structure.} \spectroviz represents critical-path behavior on a two-dimensional grid. 
The x-axis encodes the elapsed time since the start of a request \circled{1}, while the y-axis encodes the execution location \circled{2}, 
corresponding to either a service or an API, depending on the level of granularity selected by the operator. 
The lanes along the y-axis are ordered according to their most frequently observed execution order within the corresponding TPO.
Each point $(x, y)$ on the grid therefore denotes that the request was executing on the critical path at execution location $y$ at time $x$. 
When projected onto this coordinate system, an individual critical path forms a continuous trajectory that typically begins at the top-most lane, 
traverses the services participating in the request as execution progresses, and returns to the originating service upon completion. 
This representation encodes the temporal evolution of the critical path across the system components.

\fakepara{Visualizing the Aggregate Context.} \spectroviz visualizes the critical-path dimension of the baseline TPO as an occupancy probability distribution. 
Specifically, it encodes the probability that a request is executing on the critical path at execution location $y$ and time $x$ 
using a grayscale heat map \circled{3}, where white denotes a probability of 0 and black denotes a probability of 1.
The occupancy probability at each grid cell is computed by normalizing the corresponding element of the 
occupancy array in the TPO by the total number of traces represented by the TPO. 
Intuitively, darker regions, such as the one for the \codefrag{frontend} \circled{4}, correspond to portions of the execution that frequently appear on the critical path across the reference population, 
whereas lighter regions, such as the one for the \codefrag{profile\_service} \circled{5}, indicate infrequent or atypical critical-path behavior.
When the reference population consists of a single trace, each occupancy probability is either 0 or 1, 
reducing the heat map to a binary occupancy map. 
Consequently, the same visualization seamlessly supports both trace-to-population and trace-to-trace comparison without requiring a separate rendering pipeline.

\fakepara{Overlaying the Anomalous Request.} We then overlay the critical path of the anomalous
request on top of the occupancy probability distribution. 
\spectroviz renders critical-path execution segments solid line segments \circled{6} and 
transitions between services as dotted lines \circled{7} to distinguish communication from computation.
We encode the color of the critical path to be red to utilize the popout phenomenon~\cite[Ch. 5.5.4]{munzner2014vad},
making it immediately distinguishable from the underlying probability distribution.
This overlay allows regions where the anomalous execution diverges from typical behavior to naturally stand out. 
In particular, critical-path segments that pass through low-occupancy regions appear as prominent red bars against a white background,
immediately drawing the operator's attention to unlikely execution behavior. 
Moreover, the earliest such divergence along the time axis, \codefrag{search\_service} at $7$ms for this trace \circled{8}, pinpoints where the anomalous request first deviated 
from the behavior observed in the baseline trace population, providing a natural starting point for diagnosis.

\subsection{\parallax}
\label{sec:parallax}

\begin{figure}[t]%
\centering
\begin{lstlisting}
Structural Motifs:
  search.Nearby->search.RateServiceClient_GetRates: 1168
  search.Nearby->search.GeoServiceClient_Nearby: 1168
  search.RateServiceClient_GetRates->rate.GetRates: 1168
  search.GeoServiceClient_Nearby->geo.Nearby: 1168
Temporal Info:
  search.Nearby: p25: 1.665ms,p50: 2.2478ms,p75: 2.915ms,p90: 3.561ms,p99: 4.7116ms
Tags Info:
  search_success: 1168
Logs Info:
  search.Nearby:
    Search was successful: 1168
\end{lstlisting}
\caption{Excerpt of the linearized TPO representation generated by the Summarizer before being provided to the LLM for summarization and the Comparer.}%
\label{lst:tpo_str_example}
\end{figure}

\fakepara{\parallax Design.}
\parallax is comprised of two different types of LLM agents - Summarizer ($S$) and Comparer ($C$). $S$ takes as input the two TPOs that need to be compared, as well as optional context provided by the operator to focus the summarization on a specific issue. 
For each TPO, $S$ first produces a summary by flattening the TPO into
its string representation, as shown in \autoref{lst:tpo_str_example}, by translating the structural dimension of the TPO into a simple list,
converting the temporal dimension into measurements at pre-configured quantile measurements,
and the semantic information into an ordered list with the set of unique log statements.
$S$ then feeds this transformed TPO into an LLM to generate a comprehensive text summary for the TPO. 
$C$ then takes the two generated summaries as input generates the final comparison output between the two TPOs.
Note that we do not provide any insight to the model regarding what to emphasize
and what not to emphasize in the summary as well as the generated comparison.

We adopted a simple multi-agent design for \parallax as an initial implementation, 
with the intention of iteratively refining it as we identified limitations. 
In our experiments, \parallax achieved strong performance at low cost, 
suggesting that this lightweight architecture, combined with the rich information encoded in TPOs, is sufficient for effective trace comparison. 
Nevertheless, exploring more sophisticated multi-agent designs remains an important direction for future work.

Although \parallax is designed to support all three trace comparison tasks, 
it is most effective for population-to-population comparison. 
In this setting, \parallax can identify, summarize, and contrast semantic, temporal, and structural trends across trace populations. 
Such rich aggregate information is not available in a single trace, 
making \parallax less effective for pairwise trace comparison and trace-to-population comparison.
\section{Trace Comparison Benchmark}
\label{sec:benchmark}

Evaluating trace comparison systems presents a fundamental challenge. 
Existing tracing datasets typically provide traces collected from production systems, 
but offer limited control over the nature, magnitude, and location of behavioral differences. 
For example, open source trace datasets released by large cloud companies
such as Uber~\cite{lee2024tale} and Alibaba~\cite{luo2021characterizing} 
are highly anonymized and stripped with any detail information
rendering them not as good targets for evaluations and case studies. Moreover, these datasets also suffer
from incompleteness and inconsistency issues~\cite{huye2024systemizing}.
Consequently, they provide little support for systematically evaluating the efficacy of the trace comparison technique.

Existing open-source benchmarks and datasets such as the one provided by TraceLLM~\cite{zhou2026tracellm} or
by Anand et al~\cite{anand2019deathstarbenchtraces} only contain traces for pre-determined
use cases. To address this limitation of lacking diversity of traces for comparison tasks, 
we build a benchmark for comparative trace analysis, called \tracecomp
that allows users to generate populations of perturbed traces.

\fakepara{Implementation.}
We implement \tracecomp on top of Blueprint~\cite{anand2023blueprint}, a microservice implementation generator.
We choose Blueprint as the foundation for building our benchmark for three reasons.
First, Blueprint provides abstractions that decouple the specification of a system's business logic from its infrastructure and implementation details, 
enabling users to generate complete microservice implementations from high-level specifications.
\tracecomp leverages this capability to allow users to programmatically define the system behaviors they wish to elicit, 
without requiring substantial manual implementation effort.
Second, Blueprint provides a reproducible and extensible framework for executing different workloads under controlled system configurations and fault-injection scenarios. This allows users of \tracecomp to define and execute a variety of workloads and
easily collect traces representing a variety of behaviors.
Third, Blueprint's modular architecture makes it easy to extend the benchmark with new applications, system components, and behavioral perturbations. 
New behaviors can be introduced by composing existing modules or adding new system features, 
allowing \tracecomp to evolve alongside emerging distributed system architectures and trace comparison requirements

\begin{table}[t]
\centering
\begin{tabularx}{\linewidth}{llX}
\toprule
\textbf{Perturbation} & \textbf{Type} & \textbf{Exposed Knobs} \\
\midrule

Fault Injection & Structural & Affected services or APIs; Probability of injecting a fault in a request. \\
Service Slowdown & Temporal & Affected services or APIs; Range of delay to be added for each request. \\
Retry Injection & Structural & Affected services; Timeout duration, Number of retries, Retry strategy.\\
Workload Drift & Semantic & Workload Request Composition, Network Partitions. \\

\bottomrule
\end{tabularx}
\caption{Perturbation dimensions and configurable knobs supported by \tracecomp.}
\label{tab:perturbations}
\end{table}

\begin{figure}[t]
\centering%
\lstset{style=golang}
\begin{framedlisting}[language=Go,escapechar=\$]
func makePerturbation(spec wiring.WiringSpec)([]string, error) {
    services := define_services(spec)
    $\tracecompx$.ServiceSlowdownPerturbation(spec, "search_service", "100ms")
    $\tracecompx$.SpanRemovalPerturbation(spec, "user_service", 50)
    $\tracecompx$.RetryInjectionPerturbation(spec, "geo_service", 3, "100ms")
    apply_default_params(spec, services)
    return services, nil
}
\end{framedlisting}
\caption{Sample Perturbation Specification for \tracecomp}%
\label{lst:perturb}%
\end{figure}

\fakepara{Perturbation Types.} We organize \tracecomp perturbations into three broad categories corresponding to the primary dimensions 
along which distributed executions can differ: 
(i) \emph{structural} perturbations modify the execution graph by changing the set of executed spans or their dependencies; 
(ii) temporal perturbations alter the timing characteristics of an execution while largely preserving its structure; and 
(iii) semantic perturbations modify the application-level behavior of requests, such as workload composition. 
Although each perturbation is categorized according to its primary effect, 
many perturbations naturally influence multiple dimensions of execution (\eg, retry injection affects both structure and semantics).
\autoref{tab:perturbations} summarizes the perturbations currently supported by \tracecomp and
the configuration knobs exposed to users for controlling each perturbation. 
By exposing perturbations at the level of system behavior rather than trace artifacts, 
the benchmark enables users to systematically generate diverse trace populations with well-defined behavioral differences 
while retaining realistic execution characteristics.

\fakepara{Perturbation Specification.} Users specify perturbations as features in Blueprint's wiring specification, 
which defines the behavior at the infrastructure level and the overall orchestration of the generated system. 
Each perturbation is implemented either as a single Blueprint plugin or as a composition of multiple plugins that collectively realize the desired behavior. 
This modular design enables perturbations to be reused, combined, and extended to create increasingly complex execution scenarios.

For example, consider the perturbation specification shown in \autoref{lst:perturb}
for the Hotel Reservation application from DeathStarBench~\cite{gan2019open}.
The perturbation specification is defined by Lines 3--5.
Line 3 adds a service slowdown perturbation to the \codefrag{search\_service}
with a max delay of $100$ms; Line 4 adds a span removal perturbation
to the \codefrag{user\_service} with a 50\% probability of removal;
and Line 5 adds a retry injection perturbation to the \codefrag{geo\_service}
with a timeout of $100$ms and a maximum of $3$ retries on failure.
By expressing perturbations as infrastructure features, \tracecomp enables users to systematically construct a rich space of execution 
behaviors.

\fakepara{Controlled Population Generation.} 
Given a perturbation specification, \tracecomp uses Blueprint's compiler to generate and deploy the corresponding system variant. 
Users can then execute configurable workloads against the deployed system using Blueprint's workload generator, enabling the collection of trace populations under controlled workload conditions. 
During deployment, \tracecomp automatically instruments the generated services with OpenTelemetry and configures them to export spans to a Jaeger collector. 
The resulting traces are retrieved through Jaeger's API, producing a trace population corresponding to the specified perturbation and workload.
The default wiring specification with no perturbations for an application represents the baseline configuration and the traces generated by that system variant is the baseline trace population. %
\section{Evaluation}%
\label{sec:eval}

We answer the following evaluation questions:

\begin{itemize}[leftmargin=*]
    \item Can \sys enable effective trace comparative interfaces? (\autoref{sec:benchmarkres})
    \item Can \sys diagnose issues in real systems? (\autoref{sec:casestudies})
    \item What is the cost of building and using TPOs? (\autoref{sec:cost})
\end{itemize}

\subsection{\tracecomp Results}
\label{sec:benchmarkres}

In this section, we present analyses and results of using \spectroviz
and \parallax for performing the trace comparison tasks
on the trace populations generated through the \tracecomp benchmark.

\fakepara{Benchmark Setup.} We use the Hotel Reservation application from the \tracecomp benchmark to generate
a good population and a bad population of traces.
We generate the good population of the traces using the \tracecomp benchmark without adding any perturbations
to the Hotel Reservation system.
We generate various population of traces with different perturbations.
To generate each population of traces, we execute a workload that executes a mixture of the
four APIs - \codefrag{Search}, \codefrag{Recommend}, \codefrag{User}, and \codefrag{Reservation} - exported by the Frontend service in the Hotel Reservation System.

\mysubsubsection{Temporal Perturbations}

\begin{figure}[t]%
\centering%
\includegraphics[scale=\evalgraphscale]{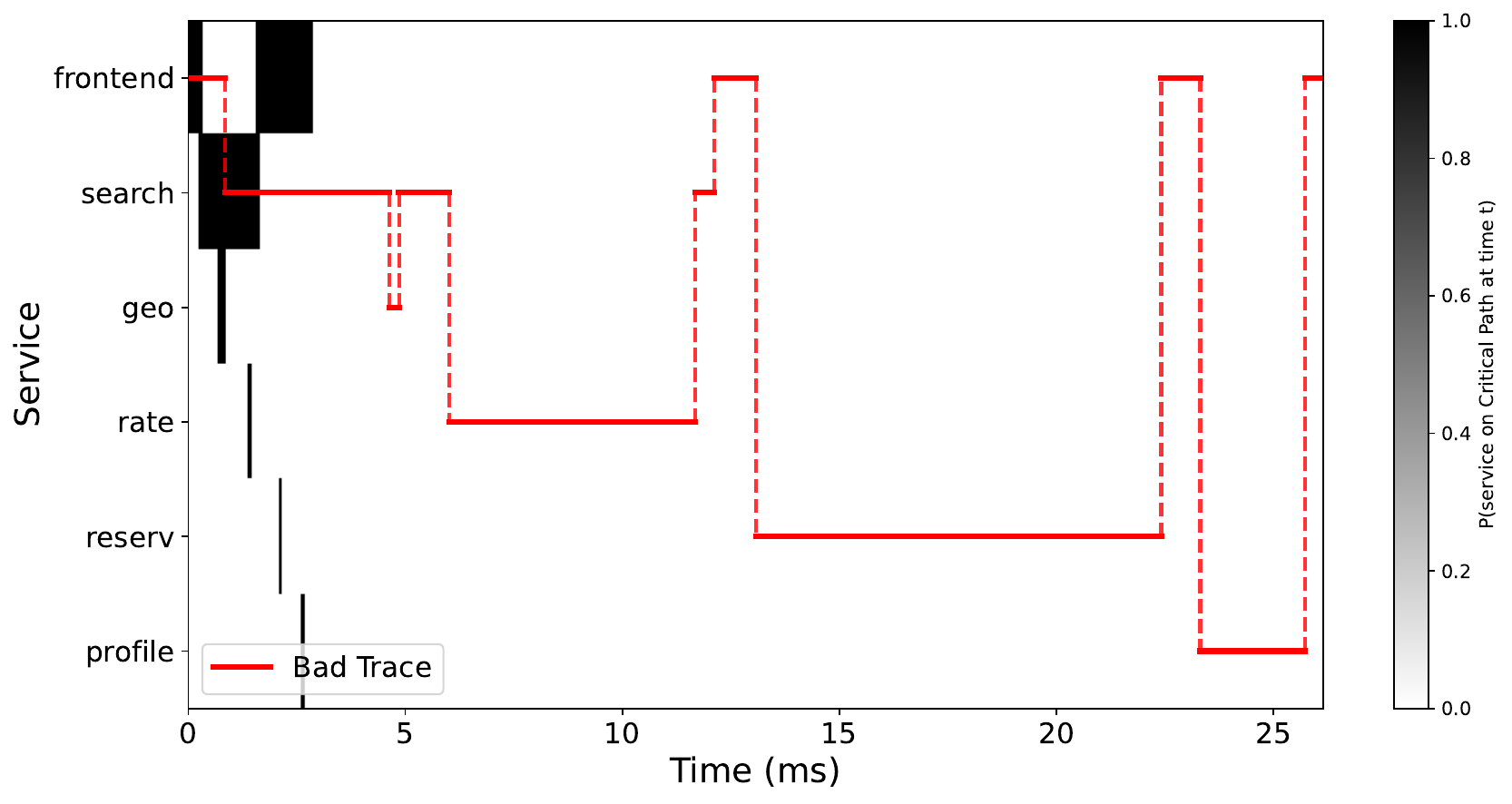}%
\caption{Pairwise trace comparison visualization generated by \spectroviz for SearchDelay.}%
\label{fig:spectroviz_temporal_diff}%
\end{figure}

\begin{figure}[t]%
\centering%
\begin{lstlisting}[escapechar=\$, postbreak={}, breaklines=true, breakatwhitespace=false, breakindent=0pt]
Overall latency is much higher in the second execution: frontend search handler latency increases from about 5.10ms p50 / 19.49ms p99 to about 13.6ms p50 / 40.0ms p99.
Search path is significantly slower in the second execution: 
frontend.Nearby rises from 3.10ms p50 / 5.99ms p99 to 9.88ms p50.  
search.Nearby rises from 2.25ms p50 / 4.71ms p99 to 8.94ms p50.  
This makes the search branch a much larger contributor in the second run.
\end{lstlisting}%
\caption{Excerpt from the population-to-population comparison generated by \parallax for SearchDelay.}%
\label{lst:parallax_temporal}
\end{figure}

\begin{figure}[t]%
\centering%
\includegraphics[scale=\evalgraphscale]{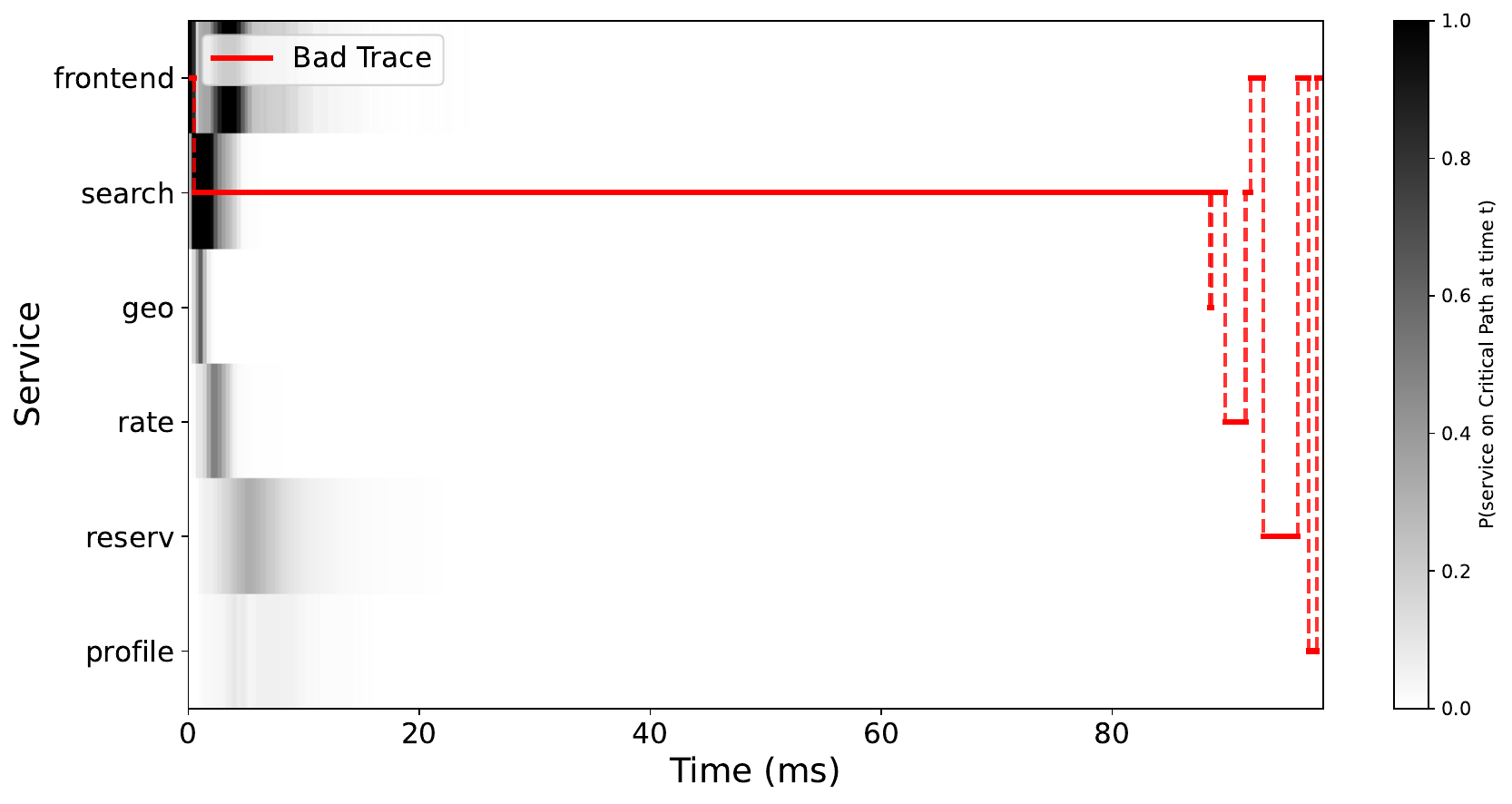}%
\caption{Trace to population comparison visualization with \spectroviz for a trace with upto 100ms injected delay.}
\label{fig:spectroviz_lat_sensitivity}
\end{figure}%

For our temporal perturbation experiment, 
we generate a bad trace population, called SearchDelay, by adding a Service Slowdown perturbation of upto $10ms$ to the \codefrag{search\_service}.
We filter both populations to only keep traces that correspond to the \codefrag{Search} API
as that is the only API that is affected by this temporal perturbation
and construct TPOs for both populations.

\fakepara{Pairwise Trace Comparison.} We sample one trace from the TPO of the bad population
of traces and compare it against a sampled trace from the TPO of the good population of traces.
\autoref{fig:spectroviz_temporal_diff} shows the visualization generated by \spectroviz
for this comparison. \spectroviz highlights the injected service delay by revealing
a clear deviation between the two traces.
The execution segment corresponding to the \codefrag{search\_service} from the bad trace is temporally elongated  
in comparison to the same segment from the good trace.

\fakepara{Trace-to-Population Comparison.} To get more insight from a comparison with the aggregate, we compare 
the selected bad trace against the good population of traces.
\autoref{fig:spectroviz_annotated} shows the visualization generated by \spectroviz for this comparison. 
Similar to the pairwise trace comparison case, \spectroviz highlights the injected temporal perturbation by revealing a clear deviation between the 
anomalous execution and the expected population behavior. 
The execution segment corresponding to the \codefrag{search\_service} from the bad trace appears temporally elongated, 
while the subsequent critical-path segments are shifted relative to the underlying occupancy probability distribution. 
This indicates that the initial slowdown in the \codefrag{search\_service} caused downstream execution stages to occur later than expected, 
allowing operators to quickly identify both the location and propagation of the performance deviation.

\fakepara{Population-to-Population Comparison.} \autoref{lst:parallax_temporal} shows an excerpt from the comparison
report generated by \parallax for the population-to-population comparison. \parallax correctly identifies
the latency issue as well as the source of the delay to be extended latency in the \codefrag{search\_service}.

\fakepara{Impact of Injected Latency on \spectroviz.} \spectroviz is highly sensitive
to temporal latency perturbations on the critical path. To emphasize the sensitivity, we generate a new bad trace population
by adding a Service Slowdown perturbation of upto $100ms$ to the \codefrag{search\_service}, we randomly sample one trace
from the bad population, and compare it against the good population of traces with \spectroviz.
\autoref{fig:spectroviz_lat_sensitivity} shows the visualization generated by \spectroviz for the comparison.
As the delay injected is higher, the execution segment corresponding to the \codefrag{search\_service} appears more elongated
and the probability distribution of the good population is restricted to the early part of the timeline.

\mysubsubsection{Structural Perturbations}

\begin{figure}[t]%
\begin{subfigure}{\linewidth}%
\centering%
\includegraphics[scale=\evalgraphscale]{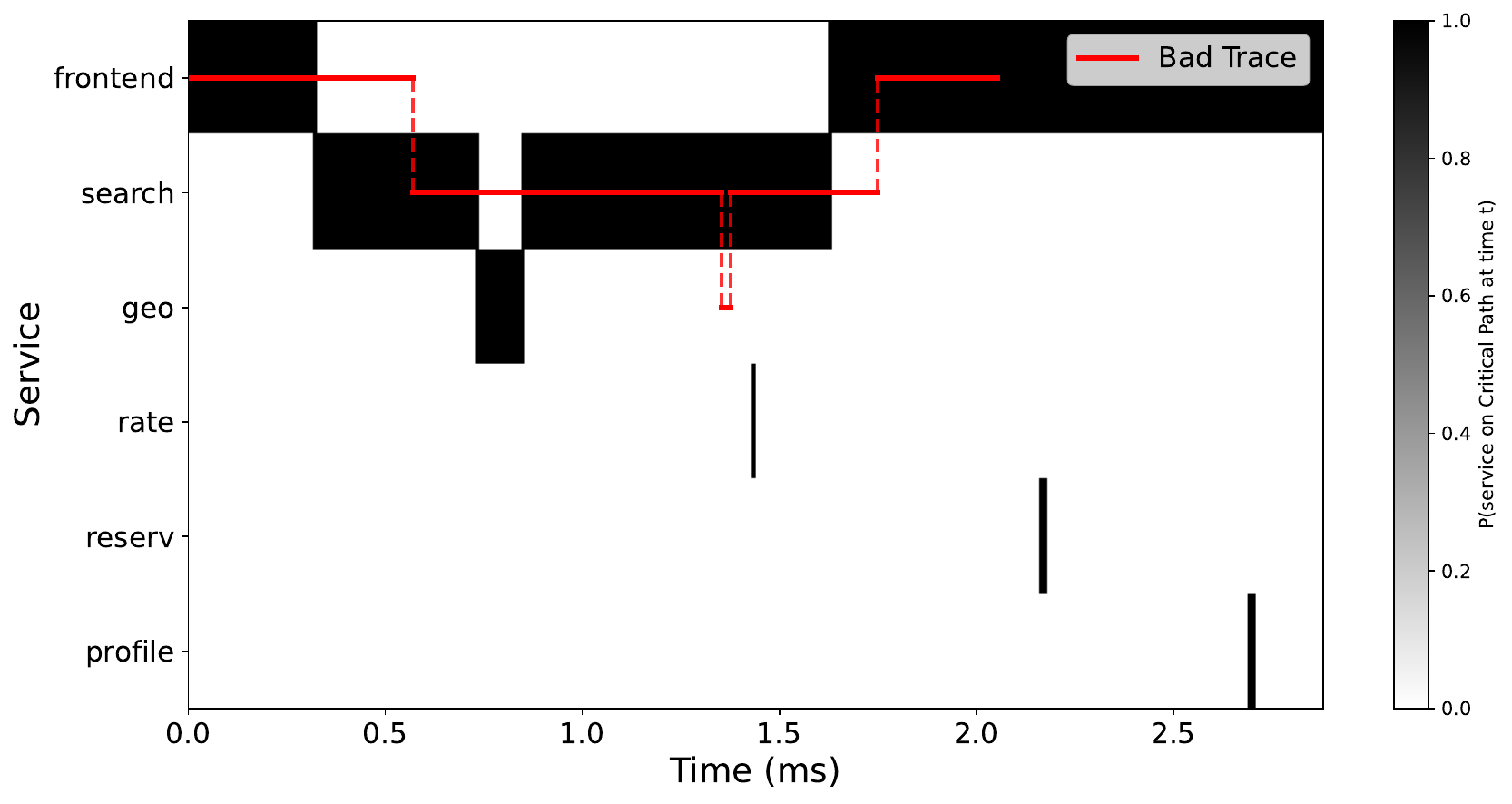}%
\caption{\spectroviz Pairwise trace comparison}%
\label{fig:spectroviz_struct_pair_case1}%
\end{subfigure}%

\begin{subfigure}{\linewidth}%
\centering%
\includegraphics[scale=\evalgraphscale]{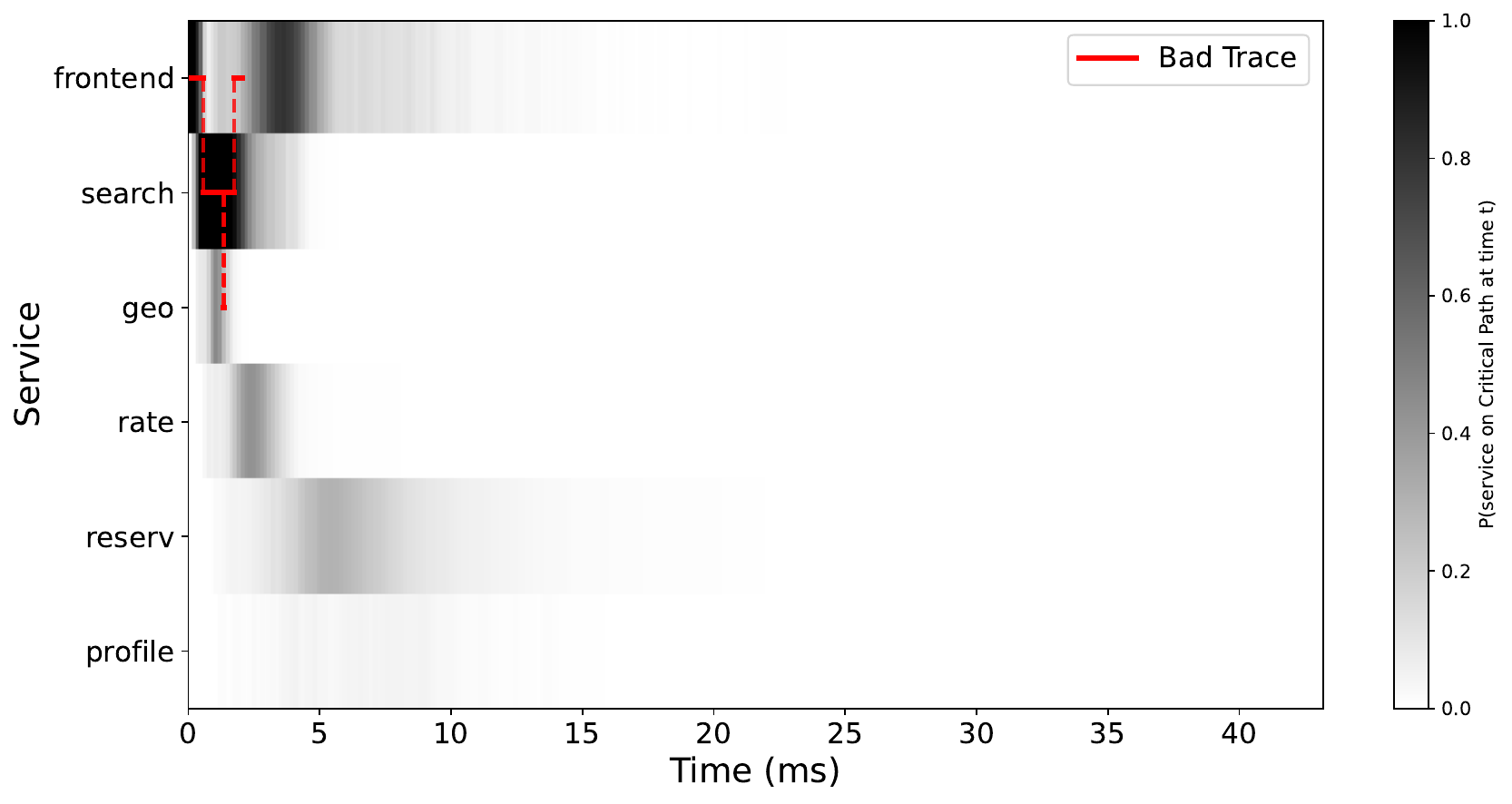}%
\caption{\spectroviz Trace to Population comparison}%
\label{fig:spectroviz_struct_agg_case1}%
\end{subfigure}%

\begin{subfigure}{\linewidth}%
\centering%
\begin{lstlisting}[escapechar=\$, postbreak={}, breaklines=true, breakatwhitespace=false, breakindent=0pt]
Second execution reflects a lighter common-case workload with optional enrichment, but it is dominated operationally by widespread injected failures on the mandatory geo/search path 
\end{lstlisting}%
\caption{\parallax Population to Population comparison}%
\label{lst:parallax_geofault}%
\end{subfigure}%
\caption{Comparisons for GeoFault trace population.}
\label{fig:spectroviz_struct_case1}
\end{figure}%

\begin{figure}[t]%
\begin{subfigure}{\linewidth}%
\centering%
\includegraphics[scale=\evalgraphscale]{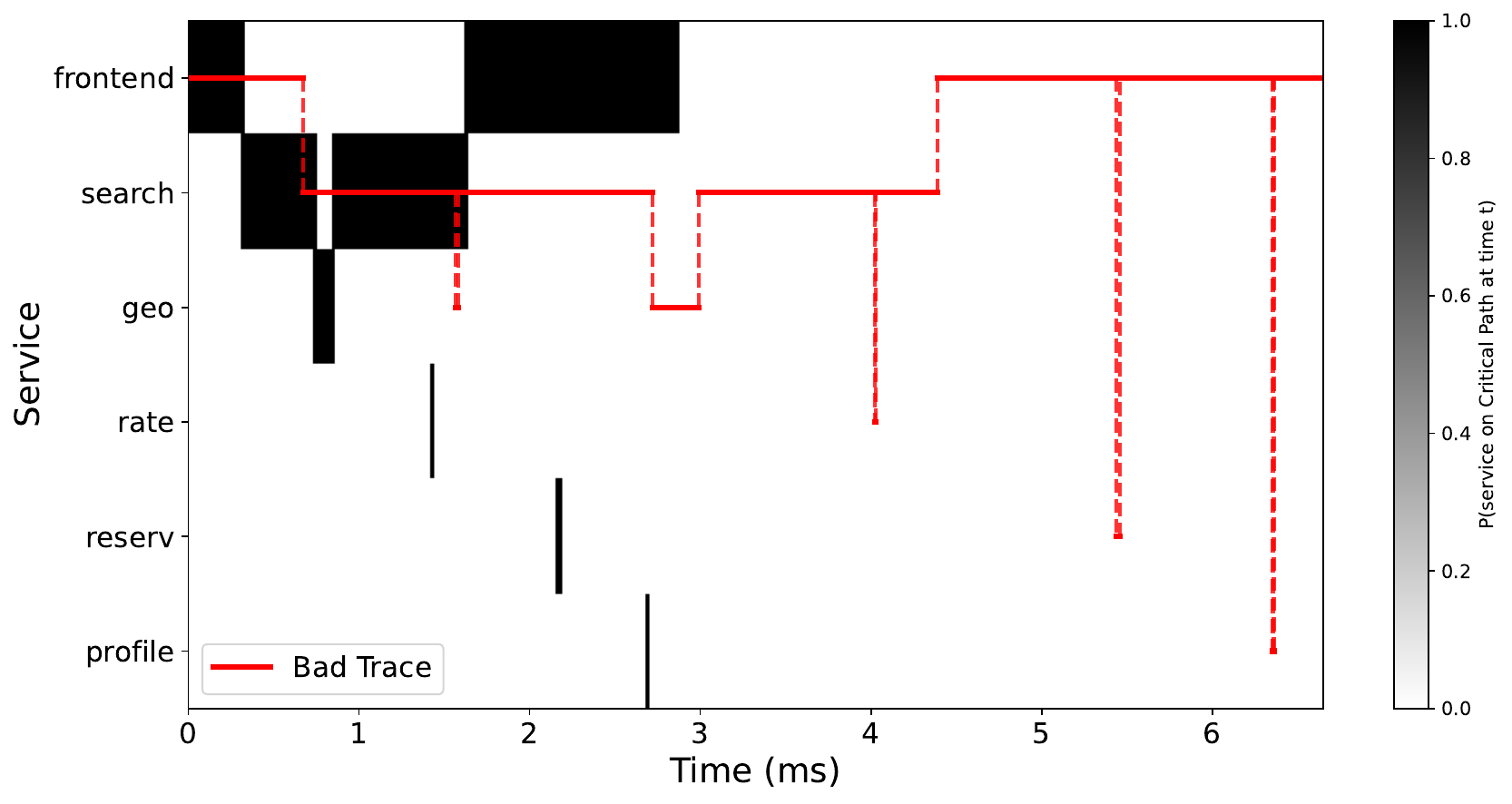}%
\caption{\spectroviz Pairwise trace comparison}%
\label{fig:spectroviz_struct_pair_case2}%
\end{subfigure}%

\begin{subfigure}{\linewidth}%
\centering%
\includegraphics[scale=\evalgraphscale]{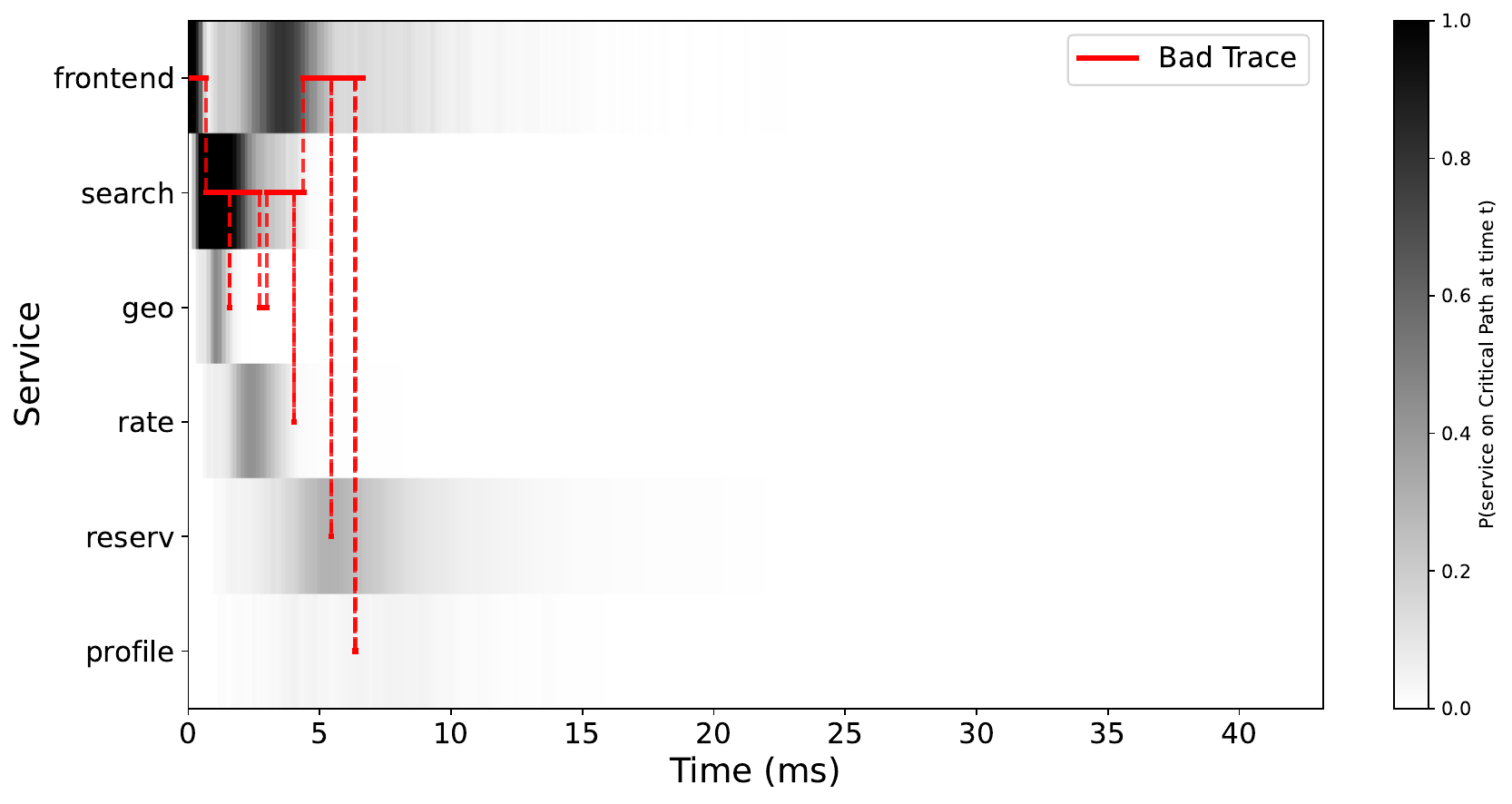}%
\caption{\spectroviz Trace to Population comparison}%
\label{fig:spectroviz_struct_agg_case2}%
\end{subfigure}%

\begin{subfigure}{\linewidth}%
\centering%
\begin{lstlisting}[escapechar=\$, postbreak={}, breaklines=true, breakatwhitespace=false, breakindent=0pt]
Geo call behavior differs: the first execution presents geo as a single stable dependency path. In the second, geo-related edges appear 1,007 times across 584 traces, indicating repeated or fan-out geo lookups within many requests.
\end{lstlisting}%
\caption{\parallax Population to Population comparison}%
\label{lst:parallax_georetry}%
\end{subfigure}%
\caption{Comparisons for GeoRetry trace population.}
\label{fig:spectroviz_struct_case2}
\end{figure}%

For our structural perturbation experiments,
we generate two bad trace populations: (i) GeoFault: population of traces generated by
adding a Fault Injection perturbation with 90\% probability to the \codefrag{geo\_service} which causes
downstream spans to not appear; and (ii) GeoRetry: population of traces generated by
adding a Fault Injection perturbation with 50\% probability and a Retry Injection with 3 retries to
the \codefrag{geo\_service}.

\fakepara{Pairwise Trace Comparison.} We sample one trace each from the TPOs of
the GeoFault and GeoRetry populations
and compare it against a sampled trace from the TPO of the good population of traces.
\autoref{fig:spectroviz_struct_pair_case1} shows the pairwise comparison
with \spectroviz for the GeoFault population. In this visualization, 
there are no execution segments for the bad trace at downstream services.
This is because the request exits early due to encountering a 
fault at \codefrag{geo\_service} and does not traverse downstream calls to other services.
\autoref{fig:spectroviz_struct_pair_case2} shows the pairwise comparison with \spectroviz
for the GeoRetry population. In this visualization, there are extra communication
calls and execution segments at the \codefrag{geo\_service} which are absent from the
good trace. This is because the bad trace suffers through a fault on its first downstream
call and then retries the execution again a second time where it succeeds.
With \spectroviz, operators can quickly identify pairwise structural differences
through either missing execution segments or through extra execution segments
and communication calls.

\fakepara{Trace-to-Population Comparison.}
We compare the selected bad traces against the good population of traces.
\autoref{fig:spectroviz_struct_agg_case1} shows the comparison with \spectroviz
for the GeoFault population. Similar to the pairwise comparison, there
are no execution segments for the bad trace at downstream services.
Moreover, it also showcases how the request execution finished much faster than
the expected execution. The lack of communication links between different
services and the temporal shift to finishing earlier serves as a strong
indicator that the request may have exited with an error.
\autoref{fig:spectroviz_struct_agg_case2} shows the comparison with \spectroviz
for the GeoRetry population. Similar to the pairwise comparison, there
are extra execution segments and communications for the bad trace at \codefrag{geo\_sevrice}.
The extra segments and communication calls capture the retry and error behavior.
With \spectroviz, operators can quickly identify ant structural irregularities
on the critical path of the bad trace.

\fakepara{Population-to-Population Comparison.} \autoref{lst:parallax_geofault} and \autoref{lst:parallax_georetry}
show excerpts generated by \parallax for the comparing the good population with the GeoFault
and GeoRetry populations. In both cases, \parallax correctly identifies the source of errors
as artifically injected failures, which correspond to the error messages in the generated traces.
For GeoRetry, \parallax also correctly identifies the multiple request execution behavior along
the path between \codefrag{search\_service} and \codefrag{geo\_service}.

\mysubsubsection{Semantic Petrubations}

\begin{figure}%
\centering
\begin{lstlisting}[escapechar=\$, postbreak={}, breaklines=true, breakatwhitespace=false, breakindent=0pt]
Second execution includes logged failures in two categories:
Rate service cache errors in 318 traces (\~54.5\%): memcache: no servers configured or available
Profile service timeout/connectivity errors in 266 traces (\~45.5\%): DeadlineExceeded ... waiting for connections to become ready
Second execution reflects a partially degraded system with a healthy core search workflow but conditional failing branches, especially profile-service timeout behavior and rate-service cache errors.
\end{lstlisting}
\caption{Excerpt from a population-to-population comparison generated by \parallax for NetFault.}
\label{lst:parallax_netfault}
\end{figure}

For our semantic perturbation experiment, we introduce a network
partition during deployment time to create a bad population of traces called NetFault. In this partition, the \codefrag{rate\_service}
is partitioned away from \codefrag{rate\_cache} and \codefrag{profile\_service}
is partitioned away from \codefrag{search\_service}. This results in majority of the
requests failing due to connectivity and timeout issues.
As semantic differences are usually only prominent across different populations,
we choose to only do population-to-population comparison for this perturbation.
\autoref{lst:parallax_netfault} shows an excerpt of the comparison report
generated by \parallax. \parallax correctly identifies the two salient semantic
differences between the populations and highlights the percentage of
traces that exhibit each type of semantic failure.

\subsection{Case Studies}
\label{sec:casestudies}

We conduct our analysis on two open source trace datasets: (i) DeathStarBenchTrace Dataset~\cite{anand2019deathstarbenchtraces}: an X-Trace~\cite{fonseca2007xtrace} trace dataset from the social network
application of the DeathStarBench microservice benchmark~\cite{gan2019open}; and (ii) an Uber Trace Dataset from CRISP~\cite{zhang2022crisp}: an open-source trace dataset released by Uber~\cite{zhang2022crisp_artifact}.

\mysubsubsection{DeathStarBench Trace Dataset}

\begin{figure}[t]%
\centering%
\begin{subfigure}{\linewidth}%
\includegraphics[scale=\evalgraphscale]{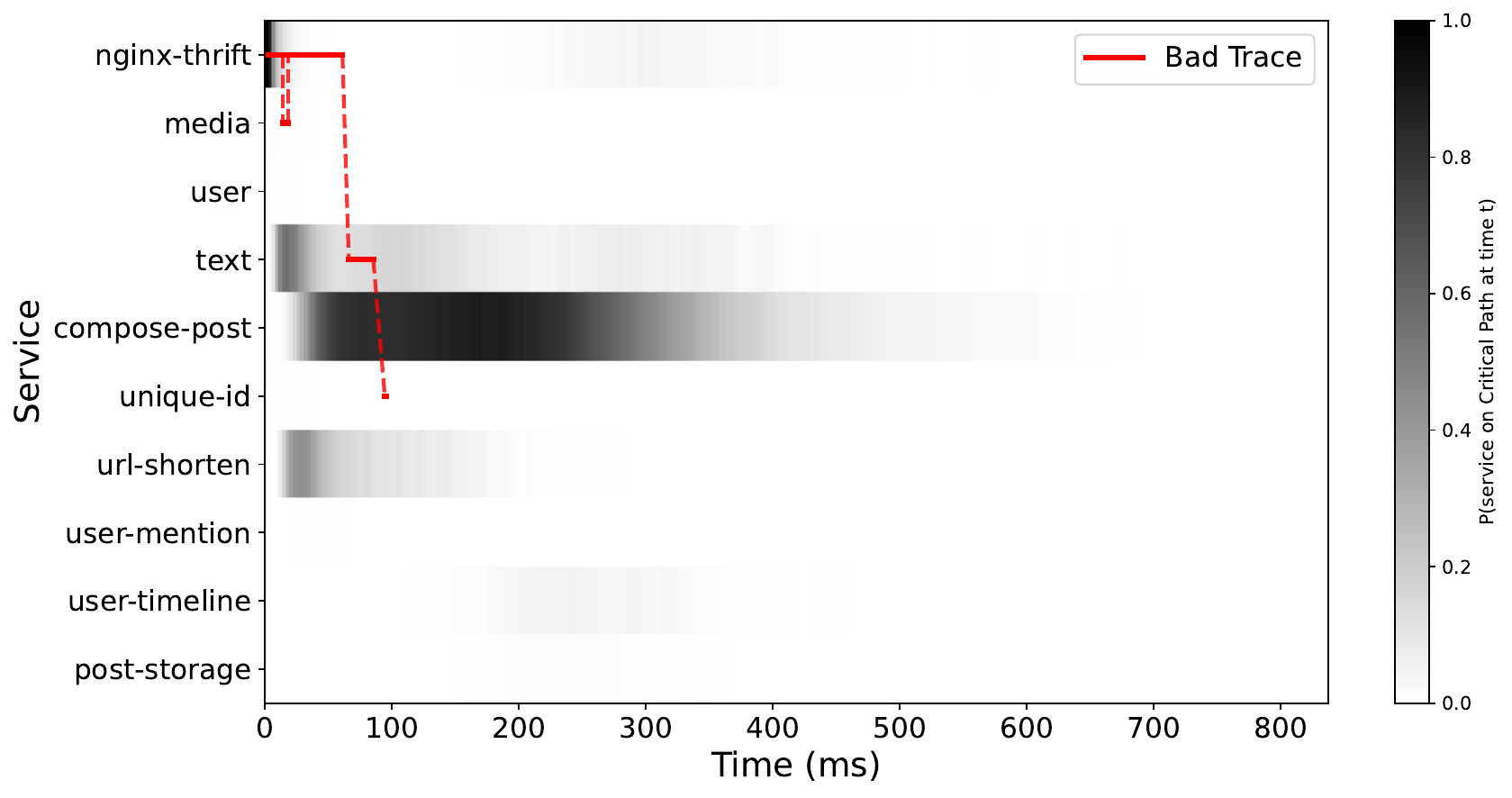}%
\caption{Trace to population visualization with \spectroviz}%
\label{fig:spectroviz_dsb_sn_broken}
\end{subfigure}

\begin{subfigure}{\linewidth}
\begin{lstlisting}[escapechar=\$, postbreak={}, breaklines=true, breakatwhitespace=false, breakindent=0pt]
$\textbf{Dominant behavior}$
First execution is dominated by a healthy fan-out/fan-in orchestration pattern. Second execution is dominated by client-pool acquisition steps, retries, and degraded alternate paths, making resource contention a first-class structural feature.
$\textbf{Latency profile}$
First execution has a slower median end-to-end latency (p50 314ms) but relatively controlled tails (p99 659ms). Second execution has a faster median (p50 54ms) but dramatically worse tail latency (p90 433ms, p99 1043ms), indicating many requests fail or short-circuit quickly while a subset stall badly.
$\textbf{Correctness / API behavior}$
First execution shows no evidence of API mismatch. Second execution uniquely shows many "Invalid method name" frontend errors, suggesting RPC interface/version/routing problems absent from the first execution.
\end{lstlisting}%
\caption{Sanitized excerpt from the Population-to-Population comparison summary generated by \parallax. Full raw summary at \autoref{lst:parallax_full_report_dsb_sn}}%
\label{lst:parallax_report_dsb_sn}
\end{subfigure}
\caption{Comparison outputs of \spectroviz and \parallax for traces from social network trace dataset~\cite{anand2019deathstarbenchtraces}}%
\label{fig:dsb_sn}
\end{figure}

The DeathStarBench Trace dataset provides two populations of \codefrag{ComposePost} API traces.
\codefrag{ComposePost} is the most complex API in the DeathStarBench Social Network application
with a request fanning out to all the services in the application. The good population
contains a set of successful requests, whereas the bad population contains
the set of traces from a time period where the system was suffering from a network partition
resulting in connectivity issues between services causing mass request failures.

We select one trace from the bad population and compare it against the good population of traces using \spectroviz.
\autoref{fig:spectroviz_dsb_sn_broken} shows the comparison of the trace with timeout issues as compared
to the good population of traces. The anomalous trace appears incomplete with missing segments across the various services
as well as is missing return and finishing segments at the entrypoint nginx. This is because 
connectivity and timeout issues cause traces to be incomplete with the root trace span exiting
without waiting for the downstream calls to exit. We validated that the critical path generated
by \spectroviz for the bad trace is same as the critical path generated by Jaeger
showing that the critical path algorithm is highly sensitive to trace completeness.

We compare the two populations of traces with \parallax.
\autoref{lst:parallax_report_dsb_sn} shows an excerpt from the population-to-population report generated by
\parallax.
\parallax correctly identifies that the differences between the two populations stem from connectivity or network partition issues, 
which manifest as request timeouts. 
It also captures the bimodal latency distribution caused by two distinct execution behaviors: early exits and long-wait requests.

\mysubsubsection{Uber Trace Dataset}

\begin{figure}[t]%
\centering%
\begin{subfigure}{\linewidth}%
\centering%
\includegraphics[scale=\evalgraphscale]{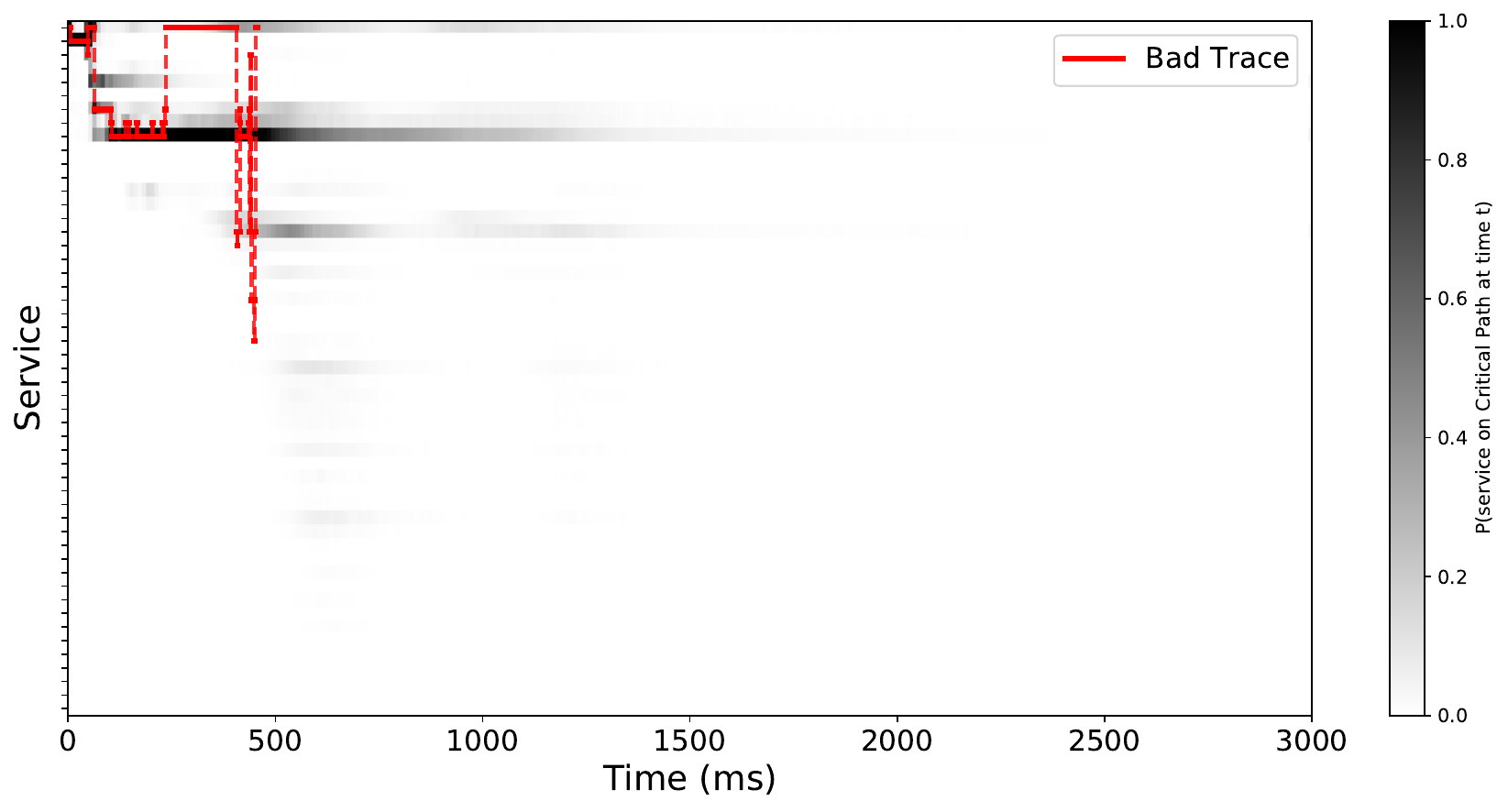}%
\caption{Trace-To-Population generated by \spectroviz}%
\label{fig:spectroviz_uber_full}
\end{subfigure}%

\begin{subfigure}{\linewidth}%
\centering%
\includegraphics[scale=\evalgraphscale]{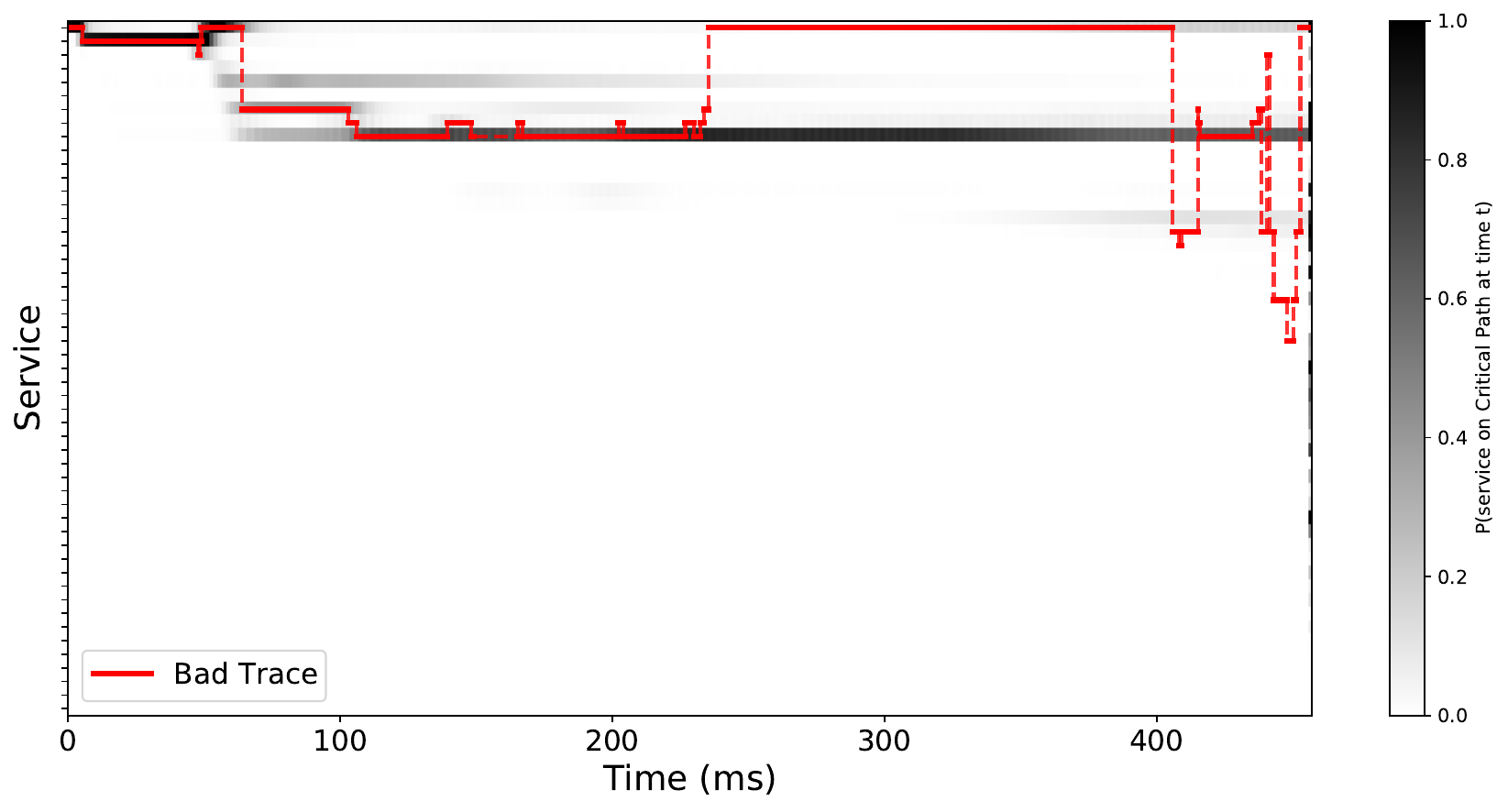}%
\caption{Zoomed-In comparison generated by \spectroviz}%
\label{fig:spectroviz_uber_full_zoom}%
\end{subfigure}%

\begin{subfigure}{\linewidth}%
\centering%
\begin{lstlisting}[escapechar=\$, postbreak={}, breaklines=true, breakatwhitespace=false, breakindent=0pt]
Canonical-path clarity is stronger in the first execution: the first describes a highly repeatable, stable backbone with downstream branches layered on top; the second describes a few dominant backbones plus a long tail of optional branches, implying relatively less structural uniformity.
\end{lstlisting}%
\caption{Excerpt from comparison report generated by \parallax}%
\label{lst:parallax_uber}%
\end{subfigure}%
\caption{Uber anomalous traces vs normal traces}%
\label{fig:uber}%
\vspace{-3mm}%
\end{figure}

The Crisp artifact~\cite{zhang2022crisp_artifact}
contains two different trace datasets: (i) \codefrag{ml-service3} which contains
$20K$ normal traces and $500$ abnormal traces that were generated
by the authors of Crisp~\cite{zhang2022crisp} by dropping 20\%
nodes in the execution graph and randomly shuffling the duration
of the nodes; and (ii) \codefrag{service43} which contains
$50K$ traces of \codefrag{operation159}.
For our case study, we use the \codefrag{ml-service3}
as it provides a population of anomalous traces.
We first find the most common root operation across both
anomalous traces and filter the populations
to only keep traces that have the same root.
This leaves a total of $14.5K$ traces in the good population
and the $349$ traces in the bad population.

We randomly sample an anomalous trace from the bad population
and compare it to the good population using \spectroviz.
\autoref{fig:spectroviz_uber_full} shows the visualization
generated by \spectroviz. As the service names are anonymized in the
trace dataset, we omit them from the generated visualization for better
readability. We notice three differences in the anomalous trace: 
(i) it exits earlier than expected;
(ii) it is missing execution segments on the critical path for the services in the bottom
half of the graph;
(iii) there are severe critical-path deviations in the $250ms$ to $400ms$ time period.
To further analyze the third difference, we use \spectroviz to create a zoomed-in visualization
that clips the comparison to the request duration of the anomalous trace as shown
in \autoref{fig:spectroviz_uber_full_zoom}. In the zoomed-in version,
we can quickly spot the deviation in that time period. Additionally, the zoomed-in version
also highlights critical-path deviations towards the end of the trace.
Next, we perform a population to population comparison using \parallax.
\autoref{lst:parallax_uber} shows an excerpt of the comparison report generated
by \parallax. As the traces are sanitized and carry almost no semantic information,
the comparison report exclusively focuses on the structural and temporal dimensions
of the TPOs. Nevertheless, \parallax still identifies the structural anomalies
in the bad population as compared to the good population.
Together, \spectroviz and \parallax are able to detect the structural and temporal
anomalies in the bad population of traces.
This shows that \spectroviz and \parallax can use TPOs to provide insight for large production traces.

\subsection{TPOs! But at what COST?}
\label{sec:cost}

\begin{table*}[t]
\begin{tabular}{lcccc}
\toprule
\textbf{System}                    & \textbf{\# Traces} & \textbf{Avg. Trace Size (KB)} & \textbf{Gen Time per Trace (s)} & \textbf{Merge Time (s)}\\
\midrule
GeoFault                           & 584  & 5.28     & 0.0004 & 0.011 \\
GeoRetry                           & 584  & 8.37     & 0.0006 & 0.023 \\
SearchDelay                        & 584  & 8.24    & 0.0006 &  0.013 \\
\tracecomp Good Pop                & 1168 & 7.42   & 0.0005 & 0.042  \\
DSB Dataset Bad                    & 1054 & 53.49    & 0.0011 &  0.069 \\
DSB Dataset Good                   & 2391 & 137.03    & 0.0030 & 0.281 \\
Uber, \codefrag{ml-service3} abnormal & 500 & 156.07 & 0.0118 & 0.698 \\
Uber, \codefrag{ml-service3} normal   & 19994 & 160.12 & 0.0132 & 37.967 \\
Uber, \codefrag{service43}            & 49999  & 235.48 & 0.0261 & 177.788 \\
\bottomrule
\end{tabular}
\caption{TPO generation and merge time for different trace datasets and their characteristics}%
\label{tab:gen_time}%
\vspace{-2mm}%
\end{table*}

\fakepara{TPO Generation Time.} \autoref{tab:gen_time} shows the average TPO generation time per trace 
and the total time required to merge TPOs across different datasets. 
The time required to generate a TPO increases with the average trace size, while the merge time grows with the number of traces being aggregated. 
Although TPO generation is inexpensive for an individual trace, it becomes the dominant cost when processing large trace collections. 
For example, for the Uber \codefrag{service43} dataset, generating TPOs for all traces using our single-threaded Python implementation takes $21$ 
minutes, whereas merging the generated TPOs takes $7\times$ less time. 
Consequently, \sys generates TPOs for each trace only once and subsequently filters and merges these TPOs according to user requirements, enabling 
repeated analyses without reprocessing the original traces.

\fakepara{LLM Cost of \parallax.} 
\parallax used minimal resources to generate the individual TPO summaries and the final comparison report for the \tracecomp comparisons.
We report \parallax token counts and cost for the \codefrag{gpt-5.2} model.
\parallax consumed $\approx2.2K$ total tokens on average to summarize a TPO
and $\approx2.3K$ total tokens on average to compare two TPOs.
For the DeathStarBench dataset,
\parallax consumed $37.47K$ input tokens and $5.69K$ output tokens which cost $\$0.18$ in total.
The significant rise in use of input and output tokens is because the dataset contains traces
that have rich event logs with a lot of semantic information.
For the Uber dataset, specifically \codefrag{Service3}, \parallax consumed a total of
$242.6K$ input tokens and $5.82K$ output tokens which cost $\$0.67$ in total.
For the filtered Uber datasets, \parallax consumed a total of $111.9K$ input tokens
and $5.21K$ output tokens which cost $\$0.36$ in total. %
\section{Related Work}%
\label{sec:related}

\fakepara{Trace Comparison Techniques.}
There exist a variety of interfaces for comparing traces~\cite{sambasivan2007categorizing,sambasivan2011diagnosing,pintrace,ekhlasi2026dtracomp,anand2020aggregate,kaldor2017canopy,dynatraceserviceflow,shkurographdiffviz,mace2013revisiting,farro2018jaegertracecompare,d2024grammar,chen2004path,barham2004using,rabo2020distributed,leone2023enhancing,traini2024vamp}
as we have discussed in \autoref{sec:tasks}.
These interfaces typically focus on specific dimensions of trace behavior and 
tailor their comparison mechanisms to effectively expose differences along those dimensions. 
We view these approaches as complementary to TPOs: rather than replacing existing comparison techniques, 
TPOs provide a dimension-preserving data model that can support diverse comparison interfaces. 
Consequently, many existing interfaces could be adapted to use TPOs as their underlying representation with minimal changes, 
allowing them to be integrated into \sys alongside \parallax and \spectroviz as complementary interfaces.

\fakepara{Visualizing Distributed System Execution.}
Visualization has long been used as a means to help operators understand the behavior of complex distributed systems 
by transforming low-level execution data into human-interpretable representations.
These techniques focus on visualizing communication patterns and execution flows
to aid general-purpose debugging~\cite{beschastnikh2020visualizing,de2000paje,de2003flexible,kunz1997poet,edwards1994dtvs,de2006execution,garduno2012theia}.
However, these systems provide visual encodings for specific diagnostic goals or dimensions of system behavior, limiting their ability to provide a unified view across diverse comparison tasks.

\fakepara{LLMs for Trace and Root Cause Analysis.}
With the rise of LLMs, there has been a recent uptick in use of LLMs and Agentic Workflows for
assisting developers and operators in root-cause analysis~\cite{roy2024exploring,xie2024cloud,zhang2026towards,zhang2026agentic,zhang2026hypothesize,las2024llexus,zhang2024lm,zhou2026tracellm,liang2026metarca,maverifyrca,tian2025gala}.
These approaches leverage the ability of LLMs to synthesize information from heterogeneous observability sources 
and potentially generate natural language explanations of system behavior.
We leverage this ability with \parallax as it can easily generate natural language
explanations of the differences between two Trace Projection Objects.
\section{Discussion}
\label{sec:discussion}

\fakepara{Integrating TPO with Intelligent Sampling.} To reduce the overhead of distributed tracing, production tracing systems typically sample only a small fraction of requests. 
Recent work has proposed adaptive sampling techniques, such as tail-based sampling~\cite{las2019sifter,las2018weighted} and retroactive sampling~\cite{zhang2023benefit}, 
that preferentially retain anomalous or otherwise interesting traces. 
These techniques are highly effective at improving the utility of collected traces under constrained storage budgets. 
However, their objective is to determine \emph{which} traces should be retained rather than \emph{why} a retained trace differs from expected behavior. 
Moreover, the rationale behind a sampling decision is typically discarded after the trace is selected. 
We believe that exposing sampling metadata—such as anomaly scores, triggering conditions, or the features that influenced the sampling decision—as 
part of the Trace Projection Object would provide valuable context for downstream comparative analysis and enable richer visualization and explanation interfaces.

\fakepara{Implications of TPO on Trace Storage.} Existing trace storage systems are primarily designed around retrieving individual traces or aggregating simple metrics, 
but \sys introduces a richer representation that supports multidimensional comparison across structural, temporal, critical-path, and semantic dimensions. 
Efficiently storing and querying TPOs requires new indexing techniques that can support dynamic population construction, 
approximate similarity search, and efficient aggregation over arbitrary subsets of traces. 
For example, an index may organize traces by structural execution patterns, critical-path occupancy signatures, or semantic attributes to accelerate the construction of comparison populations. 
Similarly, specialized storage layouts could exploit the mergeable nature of TPOs to incrementally maintain population projections as new traces arrive, avoiding repeated processing of raw traces. 
Exploring these storage abstractions could enable interactive comparative analysis over continuously evolving trace streams at production scale.
Designing indexes and storage systems optimized for TPOs is an interesting future research direction.

\fakepara{Beyond Text \& Visual Interfaces.} Typical interfaces for system analysis have relied on either textual or visual representations. 
Recent work has demonstrated the potential of auditory interfaces for analyzing execution logs~\cite{hackett2025listening} 
by leveraging humans' ability to recognize audio patterns. 
Exploring alternative modalities represents an interesting direction for future work. %
\section{Conclusions}

Comparing distributed traces is a fundamental part of the debugging process
of developers and operators. 
Existing tracing systems primarily analyze individual traces or provide aggregate analysis around
fixed dimensions, offering limited support for flexible comparison.
This paper introduced \sys, a framework that elevates trace comparison to a first-class systems abstraction. 
At the core of \sys is the Trace Projection Object (TPO), a mergeable representation that supports dynamic construction of trace populations and enables building comparison interfaces emphasizing different dimensions of shared representation. 
By decoupling representation from comparison semantics, \sys unifies multiple diagnostic interfaces while remaining grounded in representative execution traces. 
We have shown the efficacy of \sys and TPOs for both synthetic datasets generated from controlled experiments
and publicly available production datasets by Uber. %
\section*{Acknowledgments}

\fakepara{AI Usage.} ChatGPT was used primarily for editing human-written text to improve the grammar,
readability, and flow of the text. Authors also used ChatGPT as a debugging assistant
for implementing the \spectroviz interface. No agents were harmed in this process.

\if \ANON 0
\fakepara{Human Acknowledgments.} We thank Matheus Stolet, Tejas Harith, Jonas Kaufmann, and Pedro Las-Casas for providing feedback
on early drafts of the paper.
\fi 

\bibliographystyle{plain}
\bibliography{paper,bibdb/papers,bibdb/strings,bibdb/defs}

\begin{thebibliography}{10}

\bibitem{abdi2025workflow}
Mania Abdi, Peter Desnoyers, Mark Crovella, and Raja~R Sambasivan.
\newblock The workflow motif: a widely-useful performance diagnosis abstraction
  for distributed applications.
\newblock {\em arXiv preprint arXiv:2506.00749}, 2025.

\bibitem{anand2023blueprint}
Vaastav Anand, Deepak Garg, Antoine Kaufmann, and Jonathan Mace.
\newblock Blueprint: A toolchain for highly-reconfigurable microservice
  applications.
\newblock In {\em Proceedings of the 29th Symposium on Operating Systems
  Principles}, pages 482--497, 2023.

\bibitem{anand2019deathstarbenchtraces}
Vaastav Anand and Jonathan Mace.
\newblock {X-Trace trace dataset for DeathStarBench}.
\newblock Retrieved October 2019 from
  \url{https://gitlab.mpi-sws.org/cld/trace-datasets/deathstarbench\_traces/tree/master/socialNetwork},
  2019.

\bibitem{jaegertravistaissue}
Vaastav Anand and Matheus Stolet.
\newblock Jaeger issue \#6814: Feature: Augmenting trace timeline visualization
  with aggregate system statistics for faster root cause analysis.
\newblock Accessed July 2026 from
  \url{https://github.com/jaegertracing/jaeger/issues/6814}, 2025.

\bibitem{anand2020aggregate}
Vaastav Anand, Matheus Stolet, Thomas Davidson, Ivan Beschastnikh, Tamara
  Munzner, and Jonathan Mace.
\newblock Aggregate-driven trace visualizations for performance debugging.
\newblock {\em arXiv preprint arXiv:2010.13681}, 2020.

\bibitem{barham2004using}
Paul Barham, Austin Donnelly, Rebecca Isaacs, and Richard Mortier.
\newblock Using magpie for request extraction and workload modelling.
\newblock In {\em OSDI}, volume~4, pages 18--18, 2004.

\bibitem{beschastnikh2020visualizing}
Ivan Beschastnikh, Perry Liu, Albert Xing, Patty Wang, Yuriy Brun, and
  Michael~D Ernst.
\newblock Visualizing distributed system executions.
\newblock {\em ACM Transactions on Software Engineering and Methodology
  (TOSEM)}, 29(2):1--38, 2020.

\bibitem{chen2004path}
Yen-Yang~Michael Chen.
\newblock {\em Path-based failure and evolution management}.
\newblock University of California, Berkeley, 2004.

\bibitem{d2024grammar}
Andrea D'Angelo and Giordano d'Aloisio.
\newblock Grammar-based anomaly detection of microservice systems execution
  traces.
\newblock In {\em Companion of the 15th ACM/SPEC International Conference on
  Performance Engineering}, pages 77--81, 2024.

\bibitem{davidson2023qualitative}
Thomas Davidson, Emily Wall, and Jonathan Mace.
\newblock A qualitative interview study of distributed tracing visualisation: A
  characterisation of challenges and opportunities.
\newblock {\em IEEE Transactions on Visualization and Computer Graphics},
  30(7):3828--3840, 2023.

\bibitem{de2003flexible}
J~Chassin de~Kergommeaux and Benhur de~Oliveira~Stein.
\newblock Flexible performance visualization of parallel and distributed
  applications.
\newblock {\em Future Generation Computer Systems}, 19(5):735--747, 2003.

\bibitem{de2000paje}
J~Chassin De~Kergommeaux, Benhur Stein, and Pierre-Eric Bernard.
\newblock Paj{\'e}, an interactive visualization tool for tuning multi-threaded
  parallel applications.
\newblock {\em Parallel Computing}, 26(10):1253--1274, 2000.

\bibitem{de2006execution}
Wim De~Pauw, Sophia Krasikov, and John~F Morar.
\newblock Execution patterns for visualizing web services.
\newblock In {\em Proceedings of the 2006 ACM symposium on Software
  visualization}, pages 37--45, 2006.

\bibitem{dynatraceserviceflow}
Dynatrace.
\newblock Service flow.
\newblock Accessed July 2026 from
  \url{https://www.dynatrace.com/platform/service-flow/}.

\bibitem{edwards1994dtvs}
Dennis Edwards and Phil Kearns.
\newblock Dtvs: A distributed trace visualization system.
\newblock In {\em Proceedings of 1994 6th IEEE Symposium on Parallel and
  Distributed Processing}, pages 281--288. IEEE, 1994.

\bibitem{ekhlasi2026dtracomp}
Maryam Ekhlasi, Fatemeh~Faraji Daneshgar, Michel Dagenais, Maxime Lamothe,
  Naser Ezzati-Jivan, and Matthew Khouzam.
\newblock Dtracomp: Comparing distributed execution traces for understanding
  intermittent latency sources.
\newblock {\em Journal of Systems and Software}, page 112990, 2026.

\bibitem{farro2018jaegertracecompare}
Joe Farro.
\newblock Trace comparisons arrive in jaeger 1.7.
\newblock Retrieved November 2024 from
  \url{https://medium.com/jaegertracing/trace-comparisons-arrive-in-jaeger-1-7-a97ad5e2d05d},
  2018.

\bibitem{shkurographdiffviz}
Steve Flanders and Yuri Shkuro.
\newblock {A Picture is Worth a 1,000 Traces}.
\newblock Retrieved February 2020 from
  \url{https://www.shkuro.com/talks/2019-11-18-a-picture-is-worth-a-thousand-traces/}.

\bibitem{fonseca2007xtrace}
Rodrigo Fonseca, George Porter, Randy~H Katz, Scott Shenker, and Ion Stoica.
\newblock {X-Trace: A Pervasive Network Tracing Framework}.
\newblock In {\em 4th USENIX Symposium on Networked Systems Design and
  Implementation (NSDI '07)}.

\bibitem{gan2019open}
Yu~Gan, Yanqi Zhang, Dailun Cheng, Ankitha Shetty, Priyal Rathi, Nayan Katarki,
  Ariana Bruno, Justin Hu, Brian Ritchken, Brendon Jackson, et~al.
\newblock An open-source benchmark suite for microservices and their
  hardware-software implications for cloud \& edge systems.
\newblock In {\em Proceedings of the twenty-fourth international conference on
  architectural support for programming languages and operating systems}, pages
  3--18, 2019.

\bibitem{garduno2012theia}
Elmer Garduno, Soila~P Kavulya, Jiaqi Tan, Rajeev Gandhi, and Priya Narasimhan.
\newblock Theia: Visual signatures for problem diagnosis in large hadoop
  clusters.
\newblock In {\em 26th Large Installation System Administration Conference
  (LISA 12)}, pages 33--42, 2012.

\bibitem{hackett2025listening}
Finn Hackett and Ivan Beschastnikh.
\newblock Listening to the firehose: Sonifying z3’s behavior.
\newblock In {\em 2025 IEEE/ACM 47th International Conference on Software
  Engineering: New Ideas and Emerging Results (ICSE-NIER)}, pages 11--15. IEEE,
  2025.

\bibitem{pintrace}
Brittany Herr and Naoman Abbas.
\newblock {Analyzing distributed trace data}.
\newblock Retrieved March 2020 from
  \url{https://medium.com/pinterest-engineering/analyzing-distributed-trace-data-6aae58919949},
  2017.

\bibitem{huye2024systemizing}
Darby Huye, Lan Liu, and Raja~R Sambasivan.
\newblock Systemizing and mitigating topological inconsistencies in alibaba's
  microservice call-graph datasets.
\newblock In {\em Proceedings of the 15th ACM/SPEC International Conference on
  Performance Engineering}, pages 276--285, 2024.

\bibitem{huye2023lifting}
Darby Huye, Yuri Shkuro, and Raja~R Sambasivan.
\newblock Lifting the veil on $\{$Meta’s$\}$ microservice architecture:
  Analyses of topology and request workflows.
\newblock In {\em 2023 USENIX Annual Technical Conference (USENIX ATC 23)},
  pages 419--432, 2023.

\bibitem{jaeger}
{Jaeger: open source, distributed tracing platform}.
\newblock Retrieved July 2026 from \url{https://www.jaegertracing.io/}.

\bibitem{kaldor2017canopy}
Jonathan Kaldor, Jonathan Mace, Micha\l{} Bejda, Edison Gao, Wiktor Kuropatwa,
  Joe O'Neill, Kian~Win Ong, Bill Schaller, Pingjia Shan, Brendan Viscomi,
  Vinod Vekataraman, Kaushik Veeraraghavan, and Yee~Jiun Song.
\newblock {Canopy: An End-to-End Performance Tracing And Analysis System}.
\newblock In {\em 26th ACM Symposium on Operating Systems Principles (SOSP
  '17)}, 2017.

\bibitem{kunz1997poet}
Thomas Kunz, James~P. Black, David~J. Taylor, and Twan Basten.
\newblock Poet: Target-system independent visualizations of complex
  distributed-application executions.
\newblock {\em The Computer Journal}, 40(8):499--512, 1997.

\bibitem{lamport2019time}
Leslie Lamport.
\newblock Time, clocks, and the ordering of events in a distributed system.
\newblock In {\em Concurrency: the Works of Leslie Lamport}, pages 179--196.
  2019.

\bibitem{las2024llexus}
Pedro Las-Casas, Alok~Gautum Kumbhare, Rodrigo Fonseca, and Sharad Agarwal.
\newblock Llexus: an ai agent system for incident management.
\newblock {\em ACM SIGOPS Operating Systems Review}, 58(1):23--36, 2024.

\bibitem{las2018weighted}
Pedro Las-Casas, Jonathan Mace, Dorgival Guedes, and Rodrigo Fonseca.
\newblock Weighted sampling of execution traces: Capturing more needles and
  less hay.
\newblock In {\em Proceedings of the ACM Symposium on Cloud Computing}, pages
  326--332, 2018.

\bibitem{las2019sifter}
Pedro Las-Casas, Giorgi Papakerashvili, Vaastav Anand, and Jonathan Mace.
\newblock Sifter: Scalable sampling for distributed traces, without feature
  engineering.
\newblock In {\em Proceedings of the ACM Symposium on Cloud Computing}, pages
  312--324, 2019.

\bibitem{lee2024tale}
I-Ting~Angelina Lee, Zhizhou Zhang, Abhishek Parwal, and Milind Chabbi.
\newblock The tale of errors in microservices.
\newblock {\em Proceedings of the ACM on Measurement and Analysis of Computing
  Systems}, 8(3):1--36, 2024.

\bibitem{leone2023enhancing}
Jessica Leone and Luca Traini.
\newblock Enhancing trace visualizations for microservices performance
  analysis.
\newblock In {\em Companion of the 2023 ACM/SPEC International Conference on
  Performance Engineering}, pages 283--287, 2023.

\bibitem{liang2026metarca}
Shuai Liang, Pengfei Chen, Bozhe Tian, Gou Tan, Maohong Xu, Youjun Qu, Yahui
  Zhao, Yiduo Shang, and Chongkang Tan.
\newblock Metarca: A generalizable root cause analysis framework for
  cloud-native systems powered by meta causal knowledge.
\newblock {\em Proceedings of the ACM on Software Engineering},
  3(FSE):138--159, 2026.

\bibitem{lightstepcritical}
LightStep.
\newblock Service now cloud observability: View traces-seeing the critical
  path.
\newblock Retrieved July 2026 from
  \url{https://docs.lightstep.com/docs/view-traces}.

\bibitem{luo2021characterizing}
Shutian Luo, Huanle Xu, Chengzhi Lu, Kejiang Ye, Guoyao Xu, Liping Zhang,
  Yu~Ding, Jian He, and Chengzhong Xu.
\newblock Characterizing microservice dependency and performance: Alibaba trace
  analysis.
\newblock In {\em Proceedings of the ACM symposium on cloud computing}, pages
  412--426, 2021.

\bibitem{maverifyrca}
Wenrui Ma, Kang Yang, Zirui Liu, Ruili Jiang, and Yu~Jiang.
\newblock Verifyrca: Llm-guided but evidence-verified root cause analysis for
  microservice backends.

\bibitem{mace2017end}
Jonathan Mace.
\newblock End-to-end tracing: Adoption and use cases.
\newblock {\em Survey. Brown University}, 2017.

\bibitem{mace2013revisiting}
Jonathan Mace and Rodrigo Fonseca.
\newblock Revisiting end-to-end trace comparison with graph kernels.
\newblock {\em M. Sc. Project, Brown University}, 2013.

\bibitem{masson2019ddsketch}
Charles Masson, Jee~E Rim, and Homin~K Lee.
\newblock Ddsketch: A fast and fully-mergeable quantile sketch with
  relative-error guarantees.
\newblock {\em arXiv preprint arXiv:1908.10693}, 2019.

\bibitem{munzner2014vad}
Tamara Munzner.
\newblock {\em Visualization Analysis and Design}.
\newblock CRC Press, 2014.

\bibitem{opentelemetry}
OpenTelemetry.
\newblock Opentelemetry: The open standard for telemetry.
\newblock Accessed June 2026 from \url{https://opentelemetry.io/}, 2019.

\bibitem{rabo2020distributed}
Hannes Rabo.
\newblock Distributed trace comparisons for code review: A system design and
  practical evaluation, 2020.

\bibitem{roy2024exploring}
Devjeet Roy, Xuchao Zhang, Rashi Bhave, Chetan Bansal, Pedro Las-Casas, Rodrigo
  Fonseca, and Saravan Rajmohan.
\newblock Exploring llm-based agents for root cause analysis.
\newblock In {\em Companion proceedings of the 32nd ACM international
  conference on the foundations of software engineering}, pages 208--219, 2024.

\bibitem{samanta2024visualizing}
Adrita Samanta, Henry Han, Darby Huye, Lan Liu, Zhaoqi Zhang, and Raja~R
  Sambasivan.
\newblock Visualizing distributed traces in aggregate.
\newblock {\em arXiv preprint arXiv:2412.07036}, 2024.

\bibitem{sambasivan2013visualizing}
Raja~R Sambasivan, Ilari Shafer, Michelle~L Mazurek, and Gregory~R Ganger.
\newblock Visualizing request-flow comparison to aid performance diagnosis in
  distributed systems.
\newblock {\em IEEE transactions on visualization and computer graphics},
  19(12):2466--2475, 2013.

\bibitem{sambasivan2011diagnosing}
Raja~R Sambasivan, Alice~X Zheng, Michael De~Rosa, Elie Krevat, Spencer
  Whitman, Michael Stroucken, William Wang, Lianghong Xu, and Gregory~R Ganger.
\newblock Diagnosing performance changes by comparing request flows.
\newblock In {\em USENIX Conference on Networked Systems Design and
  Implementation (NSDI)}, pages 43--56. USENIX Association, March 2011.

\bibitem{sambasivan2007categorizing}
Raja~R Sambasivan, Alice~X Zheng, Eno Thereska, and Gregory~R Ganger.
\newblock Categorizing and differencing system behaviours.
\newblock {\em Hot Topics in Autonomic Computing}, page~2, 2007.

\bibitem{sigelman2010dapper}
Benjamin~H Sigelman, Luiz~Andr{\'e} Barroso, Mike Burrows, Pat Stephenson,
  Manoj Plakal, Donald Beaver, Saul Jaspan, and Chandan Shanbhag.
\newblock Dapper, a large-scale distributed systems tracing infrastructure.
\newblock 2010.

\bibitem{silva2021mu}
Sara Silva, Jaime Correia, Andre Bento, Filipe Araujo, and Raul Barbosa.
\newblock $\mu$ viz: Visualization of microservices.
\newblock In {\em 2021 25th International Conference Information Visualisation
  (IV)}, pages 120--128. IEEE, 2021.

\bibitem{sridharan2018distributed}
Cindy Sridharan.
\newblock {\em Distributed systems observability: a guide to building robust
  systems}.
\newblock O'Reilly Media, 2018.

\bibitem{sridharantraceviewwrong}
Cindy Sridharan.
\newblock Distributed tracing — we've been doing it wrong.
\newblock Retrieved February 2020 from
  \url{https://medium.com/@copyconstruct/distributed-tracing-weve-been-doing-it-wrong-39fc92a857df},
  2019.

\bibitem{tian2025gala}
Yifang Tian, Yaming Liu, Zichun Chong, Zihang Huang, and Hans-Arno Jacobsen.
\newblock Gala: Can graph-augmented large language model agentic workflows
  elevate root cause analysis?
\newblock {\em arXiv preprint arXiv:2508.12472}, 2025.

\bibitem{traini2024vamp}
Luca Traini, Jessica Leone, Giovanni Stilo, and Antinisca Di~Marco.
\newblock Vamp: Visual analytics for microservices performance.
\newblock In {\em Proceedings of the 39th ACM/SIGAPP Symposium on Applied
  Computing}, pages 1209--1218, 2024.

\bibitem{xie2024cloud}
Zhiqiang Xie, Yujia Zheng, Lizi Ottens, Kun Zhang, Christos Kozyrakis, and
  Jonathan Mace.
\newblock Cloud atlas: Efficient fault localization for cloud systems using
  language models and causal insight.
\newblock {\em arXiv preprint arXiv:2407.08694}, 2024.

\bibitem{zhang2024lm}
Dylan Zhang, Xuchao Zhang, Chetan Bansal, Pedro Las-Casas, Rodrigo Fonseca, and
  Saravan Rajmohan.
\newblock Lm-pace: Confidence estimation by large language models for effective
  root causing of cloud incidents.
\newblock In {\em Companion Proceedings of the 32nd ACM International
  Conference on the Foundations of Software Engineering}, pages 388--398, 2024.

\bibitem{zhang2023benefit}
Lei Zhang, Zhiqiang Xie, Vaastav Anand, Ymir Vigfusson, and Jonathan Mace.
\newblock The benefit of hindsight: Tracing $\{$Edge-Cases$\}$ in distributed
  systems.
\newblock In {\em 20th USENIX Symposium on Networked Systems Design and
  Implementation (NSDI 23)}, pages 321--339, 2023.

\bibitem{zhang2026towards}
Lingzhe Zhang, Tong Jia, Kangjin Wang, Chiming Duan, Minghua He, Rongqian Wang,
  Xi~Peng, Meiling Wang, Gong Zhang, Renhai Chen, et~al.
\newblock Towards in-depth root cause localization for microservices with
  multi-agent recursion-of-thought.
\newblock {\em IEEE Transactions on Dependable and Secure Computing}, 2026.

\bibitem{zhang2026agentic}
Lingzhe Zhang, Tong Jia, Yunpeng Zhai, Leyi Pan, Chiming Duan, Minghua He,
  Mengxi Jia, and Ying Li.
\newblock Agentic memory enhanced recursive reasoning for root cause
  localization in microservices.
\newblock {\em arXiv preprint arXiv:2601.02732}, 2026.

\bibitem{zhang2026hypothesize}
Lingzhe Zhang, Tong Jia, Yunpeng Zhai, Leyi Pan, Chiming Duan, Minghua He, Pei
  Xiao, and Ying Li.
\newblock Hypothesize-then-verify: Speculative root cause analysis for
  microservices with pathwise parallelism.
\newblock {\em arXiv preprint arXiv:2601.02736}, 2026.

\bibitem{zhang2022crisp}
Zhizhou Zhang, Murali~Krishna Ramanathan, Prithvi Raj, Abhishek Parwal, Timothy
  Sherwood, and Milind Chabbi.
\newblock Crisp: Critical path analysis of $\{$Large-Scale$\}$ microservice
  architectures.
\newblock In {\em 2022 USENIX Annual Technical Conference (USENIX ATC 22)},
  pages 655--672, 2022.

\bibitem{zhang2022crisp_artifact}
Zhizhou Zhang, Murali~Krishna Ramanathan, Prithvi Raj, Abhishek Parwal, Timothy
  Sherwood, and Milind Chabbi.
\newblock Crisp: Critical path analysis of large-scale microservice
  architectures (artifact), 2022.

\bibitem{zhou2026tracellm}
Tong Zhou, Xin Peng, Jie Zhang, Chaofeng Sha, Chenxi Zhang, Zicheng Yuan, and
  Senyu Xie.
\newblock Tracellm: Evaluating and exploring large language models on trace
  analysis in microservice-based web applications.
\newblock In {\em Proceedings of the ACM Web Conference 2026}, pages
  5120--5131, 2026.

\end{thebibliography}

\appendix
\section{Full \parallax Report}

\autoref{lst:parallax_full_report_dsb_sn} shows the full comparison
report generated by \parallax for comparing the two trace populations
of the \codefrag{ComposePost} API from the DeathStarBench
trace dataset~\cite{anand2019deathstarbenchtraces}.

\begin{figure*}[t]
\centering
\begin{lstlisting}
    - **Scale and consistency**
    - First execution is larger and far more uniform: **2391 traces** with a stable compose-post workflow present in almost all traces.
    - Second execution has **1054 traces** and is much less uniform, with many degraded, partial, or failed paths.
  - **Fan-out / structural completeness**
    - First execution shows near-complete fan-out from `nginx-thrift` to **user, media, text, and unique-id** in almost every trace, with strong convergence into `compose-post` and then successful downstream persistence/timeline updates.
    - Second execution shows much lower participation counts for those same branches and weaker convergence into `compose-post`, indicating many requests do not complete the normal workflow.
  - **Dominant behavior**
    - First execution is dominated by a **healthy fan-out/fan-in orchestration** pattern.
    - Second execution is dominated by **client-pool acquisition steps, retries, and degraded alternate paths**, making resource contention a first-class structural feature.
  - **Latency profile**
    - First execution has a **slower median end-to-end latency** (`nginx-thrift` p50 **314 ms**) but relatively controlled tails (p99 **659 ms**).
    - Second execution has a **faster median** (`nginx-thrift` p50 **54 ms**) but dramatically worse tail latency (p90 **433 ms**, p99 **1043 ms**), indicating many requests fail or short-circuit quickly while a subset stall badly.
  - **Main latency bottlenecks**
    - First execution's main cost is **business-path latency**, especially `text.TextHandler::UploadText` and `_ComposeAndUpload`.
    - Second execution's main cost is **resource contention and dependency waits**, especially multi-second delays in `text`/`url-shorten`/`user-mention` client-pool and upload paths.
  - **Client pool behavior**
    - First execution shows **pervasive but mostly healthy** client-pool usage, with only **rare** "No client available" / "Creating a new client" events.
    - Second execution shows **systemic pool exhaustion**, frequent waiting/timeouts, and much larger latency in pool-pop spans.
  - **Reliability / errors**
    - First execution appears **mostly successful**, with only rare cold-client events and no major correctness failures highlighted.
    - Second execution is **highly failure-prone**, with widespread connection failures, pool timeouts, and incomplete service operations.
  - **Compose-post service health**
    - In the first execution, `compose-post` acts as a stable aggregation point, reads from Redis, composes posts, persists them, and updates timelines.
    - In the second execution, `compose-post` is a **failure amplifier**, with connectivity issues, pool contention, and repeated **Redis connection failures**.
  - **Downstream persistence**
    - First execution shows `post-storage.ReadPost` and `user-timeline.WriteUserTimeline` occurring **2390 times** and remaining fast.
    - Second execution shows final compose/publish stages are **rare and unsuccessful**, with failures to connect to `post-storage` and `user-timeline`.
  - **Text and URL processing**
    - First execution's text branch is slow but functionally regular, and URL shortening is almost always used successfully.
    - Second execution's text branch is one of the most unstable parts of the system, with many failures involving `url-shorten` and degraded completion rates.
  - **Correctness / API behavior**
    - First execution shows no evidence of API mismatch.
    - Second execution uniquely shows many **"Invalid method name"** frontend errors, suggesting **RPC interface/version/routing problems** absent from the first execution.
  - **End-to-end success**
    - First execution implies a **high effective success rate** and a coherent end-to-end workflow.
    - Second execution implies a **very low effective success rate**, with only a small number of apparent successful uploads and even those encountering failures in final publication stages.
\end{lstlisting}
\caption{Full report generated by \parallax for the population to population comparison of the DeathStarBench traces}
\label{lst:parallax_full_report_dsb_sn}
\end{figure*} 
\label{page:last}
\end{document}